\newif\ifOneColumn

\ifOneColumn
    \documentclass[journal,12pt,onecolumn,draftclsnofoot,]{IEEEtran}
\else
    \documentclass[journal,twocolumn]{IEEEtran}
\fi

\usepackage{amsmath,graphicx,algorithmic,float,cite,dsfont,amssymb}
\usepackage{amsfonts,multirow,bm,array,setspace} 
\usepackage{textcomp}
\usepackage{subfigure}
\usepackage{psfrag}
\usepackage{color}
\usepackage[linesnumbered,ruled]{algorithm2e}
\usepackage{pifont,enumitem}

\newcommand\topstrut{\rule{0pt}{2.2ex}} 

\usepackage[margin=0.87in]{geometry}

\ifCLASSINFOpdf
\else
\fi

\begin{document}

\title{Scalable Deep Reinforcement Learning for Routing and Spectrum Access in Physical Layer}

\author{\IEEEauthorblockN{Wei Cui, \IEEEmembership{Student Member,~IEEE}, and Wei Yu, \IEEEmembership{Fellow,~IEEE}} 
\thanks{Manuscript submitted on March 31, 2021, revised July 14, 2021 and September 10, 2021. The materials in this paper have been presented in part at the IEEE International Conference on Acoustics, Speech, and Signal Processing (ICASSP),
Toronto, Canada, May 2021 \cite{icassp}. This work is supported by Natural Sciences and Engineering Research Council (NSERC) of Canada.  
The authors are with The
Edward S.~Rogers Sr.~Department of Electrical and Computer Engineering,
University of Toronto, Toronto, ON M5S 3G4, Canada 
(e-mails: \{cuiwei2, weiyu\}@ece.utoronto.ca).}
}


\maketitle

\begin{abstract}
This paper proposes a novel scalable reinforcement learning approach for simultaneous routing and spectrum access in wireless ad-hoc networks. In most previous works on reinforcement learning for network optimization, the network topology is assumed to be fixed, and a different agent is trained for each transmission node---this limits scalability and generalizability. Further, routing and spectrum access are typically treated as separate tasks. Moreover, the optimization objective is usually a cumulative metric along the
route, e.g., number of hops or delay. In this paper, we account for the physical-layer
signal-to-interference-plus-noise ratio (SINR) in a wireless network and further show that \emph{bottleneck} objective such as the minimum SINR along the route can also be optimized effectively using reinforcement learning. Specifically, we propose a scalable approach in which a single agent is associated with
each flow and makes routing and spectrum access decisions as it moves along the frontier nodes. The agent is trained according to the physical-layer characteristics of the environment using a novel rewarding scheme based on the Monte Carlo estimation of the future bottleneck SINR. It learns to avoid interference by intelligently making joint routing and spectrum allocation decisions based on the geographical location information of the neighbouring nodes.
\end{abstract}


\IEEEpeerreviewmaketitle







\section{Introduction}
Routing in wireless ad-hoc networks is a complex problem involving sequential
decisions in each hop in order to build a route that optimizes certain network
objective. Due to the lack of centralized control, routing in wireless ad-hoc
networks needs to be performed in a distributed manner. Starting from the
source node, the selection of next node in each hop requires considerations of
multiple often conflicting factors. This paper advocates the use of
reinforcement learning for solving this complex problem. 

This paper differs from myriad prior works in both the traditional routing
literature and in more recent works using machine learning for network
optimization in several important aspects. First, for wireless ad-hoc networks
in which nodes operate in a shared wireless medium, it is important to consider
the physical-layer characteristics of the transmissions. Classic works on
wireless ad-hoc network routing protocols, such as \cite{DSDV,aodv,tora}, often
abstract away the physical layer: the routing table is built based on a fixed
topology; the nodes do not have complete flexibility to connect to every other node. Furthermore, the mutual interference between the
transmission links and more importantly the abilities for each link to utilize
different spectrum slots are not taken into account. Although more recent works
\cite{ssa,aqor,ssbr} consider signal qualities, they utilize only heuristic
rules and do not benefit from detailed physical-layer characterizations in term
of the signal-to-interference-plus-noise ratio (SINR). The aim of this paper is
to provide a solution to the joint routing and spectrum access problem in a
large-scale network, while taking into account the physical-layer channel
strengths and the interference levels in different spectrum slots.  As these
factors are largely determined by the geographical locations of the mobile
nodes (assuming a path-loss model for wireless propagation), they motivate us
to propose a new approach to routing based on the locations of each node and
its neighbours and the physical-layer characteristics of the surrounding
wireless environment. 

This paper also explores new network objective and
new methodologies for optimizing routing. As routing amounts to successive
selections of nodes and spectrum slots in each hop, the number of possibilities
in the overall optimization problem is combinatorial in nature; therefore
finding the globally optimal route by exhaustive search is computationally
prohibitive in large networks. Decentralized discrete optimization algorithms
previously proposed for routing \cite{gafni, DSDV, aodv, tora, ssa, aqor, ssbr,
murthy, spine, shiguang} are mostly heuristic in nature. On the other hand,
routing is also a sequential decision making problem, in which each decision is
a function of the current state of the network. Thus, routing can be modeled as
a Markov decision process \cite{markov}, which naturally fits into the realm of
reinforcement learning.  In this direction, many previous works \cite{qrouting,
choi, chang, forster, santhi, alharbi, varun, leonardo, rahul, nurmi} have
employed the classical \emph{Q-learning} \cite{qlearning} or deep reinforcement
learning \cite{DQN} to train agents to find the optimal route; see
\cite{DRL_survey} for a comprehensive survey of the applications of
reinforcement learning to communications and networking.

The existing works on using reinforcement learning for routing, however, tend
to suffer from two issues. First, in most of these works, a distinct agent is
associated with each transmission node. Each agent maintains a list of
reachable neighbors and a $Q$-value table of their ``fitness scores" as the potential next hop (e.g., time or number of hops to reach the
destination). After training, the route is collectively determined by all the
agents, each selecting the neighbor with the highest fitness scores as the next hop.  
However, training a distinct agent for each node limits scalability.

Secondly, most existing works assume that the network objective is cumulative 
along the routes, e.g., as in minimizing the transmission delay or the 
number of hops. 
While cumulative rewards fit naturally in the reinforcement learning framework, 
there are also important network metrics that are not cumulative in nature. 
In particular, assuming adaptive coding and modulation in each link, the
maximum overall throughput that can be supported in a route is the minimum rate
across all the hops, so it is a function of the \emph{bottleneck} SINR within
the data flow. Such bottleneck objective does not easily fit in the Markov 
decision process framework. A main objective of this paper is to show how to
train a reinforcement learning agent to maximize the bottleneck objective.

To summarize, the approaches in most of the existing works in using
reinforcement learning on routing for ad-hoc wireless networks are restricted
by all or at least some of the following modeling and design choices:
\begin{enumerate}[label=\textbf{C.\arabic*}]
    \item The network model assumes a fixed set of connections, where two nodes are either reachable or not.\label{enum:2}
    \item A distinct agent is associated with every node in the network and is trained specifically for this node.\label{enum:1}
    \item The objective is cumulative along the route. \label{enum:4}
    \item The same network topology must be maintained from training to testing.  \label{enum:3}
\end{enumerate}
Each of these modeling and design choices imposes limitations and has drawbacks. 
\ref{enum:2} abstracts away pivotal physical-layer characteristics of the network, thus preventing optimization of objectives such as signal-to-noise-and-interference (SINR) based quality-of-service (QoS). 
\ref{enum:1} requires the number of agents to scale as the network size, therefore limiting scalability and generalization capability. 
\ref{enum:4} does not allow the optimization of objectives such as the bottleneck link rate within a data flow. 
\ref{enum:3} further limits agents' abilities to adapt to network changes. 
Attempting to address the limitation of \ref{enum:3}, \cite{chang} let agents
\emph{explore} all new neighbors opportunistically, which is essentially
retraining.  But as the layout of an ad-hoc network can frequently change (e.g., due to mobility), any
solution designed under \ref{enum:3} is inherently insufficient. Further, to
address the limitation of \ref{enum:1}, \cite{lspirouting} proposes the use of
a specific $Q$-function estimator \cite{lspi} for the agent, allowing the same
agent to be applied to different nodes. But, none of these works perform
routing based on physical-layer attributes. 

In this work, we recognize the critical importance of making routing decisions
based on the physical-layer characteristics of the network, and moreover the
importance of optimizing spectrum access jointly with routing decisions. This is
because the allocation of which frequency bands to transmit for a particular
link significantly affects the interference levels of all nearby links,
including the previous and subsequent links in the same flow. While spectrum
access has been the focus in many previous work in wireless network optimization, 
including conventional optimization algorithms \cite{dual,yiping,songgao}, 
supervised learning \cite{peiliang}, and reinforcement learning \cite{qingzhao,mosleh,yiding,yingchang1,yingchang2}, 
all these works assume fixed wireless connections, thus do not make routing
decision jointly with spectrum optimization. While there exist
conventional optimization based works that focus on both routing and spectrum
access \cite{chunsheng,yongk,talay}, these works do not take advantage of
learning based method. We mention here that \cite{sudeep} explores
learning automata based optimization over wireless sensor networks that
includes both routing and spectrum access, but its focus is exclusively on
energy conservation.

The main contribution of this paper is a novel and scalable deep reinforcement
learning approach to simultaneous routing and spectrum access for ad-hoc
wireless networks based on physical-layer inputs, constraints, and objectives.
We focus on the maximization of the bottleneck rate across the flow as the main
task, and solve the spectrum access problem along the routing process by
training a reinforcement agent to determine the suitabilities of different
frequency bands for each link. For routing, as a main novelty, we associate one
agent to each \emph{data flow} from the source node to the destination.  The
agent executes sequential node-selection tasks along the route.  Furthermore,
we propose a novel \emph{Monte-Carlo} based reward estimation scheme within the
conventional Q-learning strategy \cite{qlearning} to enable the agent to
effectively learn to optimize a bottleneck objective. Throughout the paper, we propose to model the network environment via a local node density function. This is in line with the physical-layer modeling of the ad-hoc network, and more importantly, also aligns with the reinforcement learning framework which requires the environment to be stationary.
 
The fact that the agent takes physical-layer information as input is
crucial---this allows the same agent to be used for all nodes along the route,
thus providing scalability.  We show that the same model parameters for the
agent can also be shared across distinct data flows, or even generalized across
different ad-hoc networks.
Finally, the agent can also be designed to adapt to network characteristics
such as the density of neighbors.  

The rest of the paper is organized as follows. Section~\ref{sec:problem}
establishes the system model and formulates the optimization problem.
Section~\ref{sec:method} proposes a deep reinforcement learning based approach
for routing and spectrum access in wireless ad-hoc networks. The performance of
the proposed method is analyzed in Section~\ref{sec:results}. As extensions,
Section~\ref{sec:delay} incorporates delay into the objective;
Section~\ref{sec:powercontrol} performs power control to further improve the
data rate.  Finally, conclusions are drawn in Section~\ref{sec:conclusion}.

\section{Routing in Ad-hoc Wireless Network}\label{sec:problem}

\subsection{Network Model}

Consider a wireless ad-hoc network supporting $F$ data flows, each with a fixed
source and a fixed destination node, and routed through multiple hops over 
a subset of $N$ intermediary relay nodes in the network. We assume that there
are $B$ available frequency bands each of bandwidth $W$ Hz, 
and each hop uses one of the $B$ frequency bands for signal transmission and reception. We use $\mathcal{F}$ to
denote the set of data flows; $\mathcal{S}$ and $\mathcal{T}$ to denote the set
of sources and destinations, respectively; $\mathcal{N}$ to denote the set
of relay nodes; and $\mathcal{B}$ to denote the set of available frequency
bands, with $|\mathcal{F}|=F$, $|\mathcal{N}|=N$, and $|\mathcal{B}|=B$. 

We assume that the environment around each node follows some statistical distribution. To this end, we characterize the ad-hoc network through a node density distribution function, and assume some local density of the neighboring
nodes across each sub-region. Throughout this paper, during routing in a given network topology, the locations of the nodes are assumed to be fixed. Routing in the presence of mobility would be a much
more challenging problem. 

\subsection{Routing and Spectrum Access}

The task of routing is to select an ordered list of relay nodes 
for each flow $f\in\mathcal{F}$. The task of spectrum access is to select a
frequency band in $\mathcal{B}$ for each hop. Mathematically, we
represent the route for flow $f$ as an ordered list denoted by
$\mathbf{n}^{(f)}$: 
\begin{align}\label{equ:nodelist}
	\mathbf{n}^{(f)}=(n^{(f)}_0, n^{(f)}_1, n^{(f)}_2\dots n^{(f)}_h, n^{(f)}_{h+1})
\end{align}
where $n^{(f)}_0 = s_f\in\mathcal{S}$ represents the source node for $f$,
$ n^{(f)}_{h+1}=t_f\in\mathcal{T}$ represents the destination node for $f$, and $n^{(f)}_i \in\mathcal{N}, i=1,\cdots,h$ are the relay nodes. We represent the spectrum
access for flow $f$ as an ordered list denoted by $\mathbf{b}^{(f)}$, containing
the frequency bands of each hop 
\begin{align}\label{equ:bandlist}
    \mathbf{b}^{(f)}=(b^{(f)}_0, b^{(f)}_1, b^{(f)}_2\dots b^{(f)}_h)
\end{align}
where $b^{(f)}_j \in\mathcal{B}, j=0,\cdots,h$. 
Note that each hop uses only one frequency band with a fixed bandwidth.

Given $\mathbf{n}^{(f)}$ and $\mathbf{b}^{(f)}$, we have a complete description of the data flow $f$ as follows:
\begin{align}\label{equ:dataflow}
    n^{(f)}_0\xrightarrow{b^{(f)}_0}n^{(f)}_1\xrightarrow{b^{(f)}_1}n^{(f)}_2\xrightarrow{b^{(f)}_2}\dots\to n^{(f)}_h\xrightarrow{b^{(f)}_h}n^{(f)}_{h+1}
\end{align}
In establishing $f$, there are $h$ routing decisions for selecting the nodes 
for the $h$ hops, as well as $h+1$ spectrum access decisions for selecting the $h+1$ transmission bands. Collectively, these routing and spectrum access decisions over all the flows are denoted as:
\begin{subequations}\label{equ:solset}
\begin{align}
	\mathbf{N}_{\mathcal F} =  \{\mathbf{n}^{(f)}, \ \forall f\in\mathcal{F}\}, \\
	\mathbf{B}_{\mathcal F} =  \{\mathbf{b}^{(f)}, \ \forall f\in\mathcal{F}\}. 
\end{align}
\end{subequations}

We restrict each hop of any data flow to use only one frequency band, which is already implicitly reflected in (\ref{equ:dataflow}).
We do not allow a node to transmit and receive in the same frequency band, but 
we do allow for the possibility that an intermediary node can serve as a relay for 
multiple data flows, as long as all of its incoming and outgoing
hops occupy distinct frequency bands. 
Mathematically, this means for a flow $f$ with $h$ relays
\begin{align}\label{equ:sameflowband}
	b^{(f)}_i \neq b^{(f)}_{i+1}, \quad \forall i = 0, 1, \cdots, h-1;
\end{align}
and if $n_i^{(f_1)} = n_j^{(f_2)}$ for some $f_1, f_2 \in \mathcal{F}$ and some $i, j$, then 
\begin{align}\label{equ:diffflowband}
    \{ b^{(f_1)}_{i-1}, b^{(f_1)}_i, b^{(f_2)}_{j-1}, b^{(f_2)}_j \} \quad\text{are all distinct.}
\end{align} 
Note that (\ref{equ:diffflowband}) subsumes (\ref{equ:sameflowband}).

Finally, to eliminate loops in routing (which is never helpful), we explicitly 
enforce that each data flow can visit an intermediary node at most once 
\begin{align}
    n^{(f)}_i\neq n^{(f)}_j, \quad \forall i\neq j.
\end{align}

\subsection{Performance Metrics}\label{sec:physicalobjectives}

At the physical layer, the maximum transmission rate between a pair of nodes
over a frequency band is characterized by the capacity of the underlying
wireless channel, which is a function of the signal-to-noise-and-interference
(SINR) at the receiver. 
In details, consider a link with transmitting node $i\in \mathcal{S} \cup \mathcal{N}$ 
and receiving node $j \in \mathcal{T} \cup \mathcal{N}$, over
the frequency band $b \in \mathcal{B}$ with bandwidth $W$, with $p_i$ as the
transmit power of node $i$ and $\sigma^2$ as the background noise power in each frequency band, and 
$h_{ij,b}\in\mathcal{C}$ as the channel strength from node $i$ to node $j$ 
in the band $b$, the maximum transmission rate of this link is related to 
the SINR as follows:
\begin{subequations}\label{equ:rate}
\begin{align} 
\text{SINR}_{(i,j,b)} =& \frac{|h_{ij,b}|^2p_ix_{i,b}}{\sum_{\substack{k\neq i,j \\ k\in \mathcal{S} \cup \mathcal{N}}}|h_{kj,b}|^2 p_kx_{k,b} + \sigma^2} \label{equ:SINR} \\
R_{(i,j,b)} =& W\log\left(1+\text{SINR}_{(i,j,b)}\right) \label{equ:instantRate}
\end{align}
\end{subequations}
The binary variable $x_{i,b}\:(i\in\mathcal{S}\cup\mathcal{N}, b\in\mathcal{B})$ indicates whether the node $i$ is transmitting on the band $b$ or idle, and is determined by the routing and spectrum access solution (i.e. if node $i$ is used by any route to transmit over the band $b$ or not), i.e.,
\begin{align}
    x_{i,b}=
    \begin{cases}
	    1 \quad\text{if}\quad\exists f,k: n^{(f)}_k=i {\ \rm and\ } b^{(f)}_k=b \\
    0 \quad\text{otherwise.}
    \end{cases}
\end{align}

\begin{figure}
    \centering
    \ifOneColumn
        \includegraphics[width=0.55\textwidth]{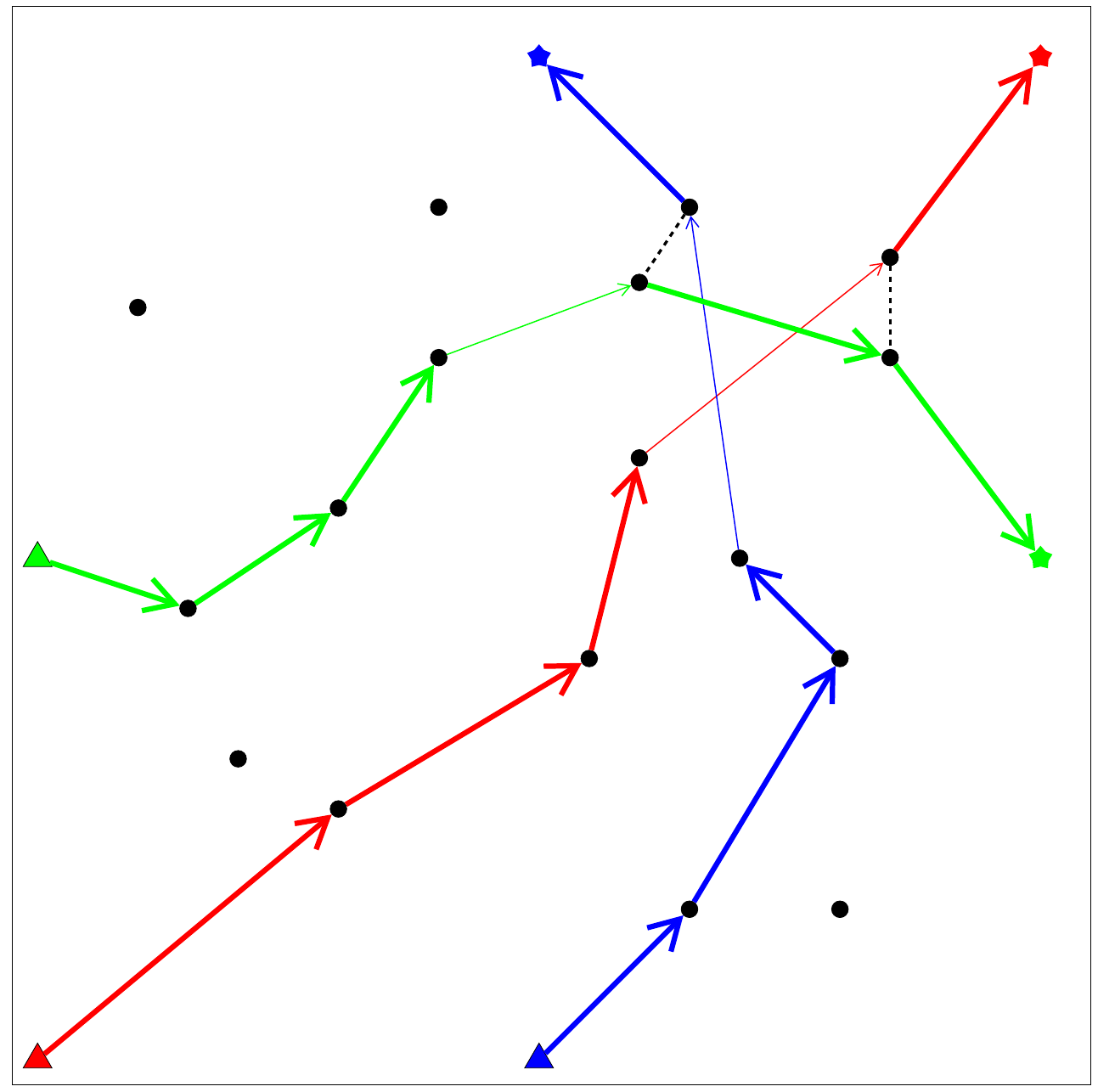}
    \else
        \includegraphics[width=0.4\textwidth]{Figures/Figure_Routing_Bottleneck.pdf}
    \fi
    \caption{Bottleneck links (marked with thin arrows) due to strong interference (marked with dotted lines) from nearby links}
    \label{fig:bottleneck}
\end{figure}

We assume full-buffered transmission in which the data is continuously
transmitted along the route. In this case, the overall data rate for each
flow is determined by its \emph{bottleneck} link capacity (i.e. the
minimum data rate among all its links). Specifically, consider the 
flow $f$ with its routing and spectrum access specified as in (\ref{equ:dataflow}),
the overall data rate of the flow $f$ is
\begin{align}\label{equ:bottleneckrate}
	R^{(f)} = \min_{i=0,1,2,\dots,h}R_{(n^{(f)}_i, n^{(f)}_{i+1}, b^{(f)}_i)}.
\end{align}
The concept of the bottleneck link is visually illustrated in Fig.~\ref{fig:bottleneck}. 
We denote the set of the overall data rates of all flows to be 
\begin{align}\label{equ:rateset}
	\mathbf{R}_{\mathcal F} = \{R^{(f)}, \ \forall f\in\mathcal{F}\}.
\end{align}

\subsection{Optimization Problem Formulation}\label{sec:problemformulation}

The objective for optimal routing and spectrum access is to determine the
set of solutions as in (\ref{equ:dataflow}) for each data flow in order to 
optimize some global objective across all the flows in the network. The
global objective is called network utility, denoted as $U(\cdot)$, which 
is a function of the data rates of all the flows (e.g. the sum data rates 
or the minimum data rate). 
Mathematically, the optimization problem is formulated as follows:
\begin{subequations}\label{equ:problem}
\begin{align}
	\underset{{\mathbf N}_\mathcal{F},{\mathbf B}_\mathcal{F}} {\text{maximize}}\quad &  U({\mathbf R}_\mathcal{F}) \\
    \text{subject to}\quad& (\ref{equ:nodelist}),(\ref{equ:bandlist}),(\ref{equ:solset})-(\ref{equ:rateset}).
\end{align}
\end{subequations}
Note that the data rates of different links both within each flow and across
the flows have strong inter-dependencies due to the mutual interference.
Specifically, if two links share the same frequency band, then the transmitted
signal of one link appears as interference for the other link. Thus, the
spectrum allocation problem is tightly coupled with the routing problem.
Consequently, there is a trade-off between the data rates of different flows in
the network.

We emphasize that (\ref{equ:problem}) is an integer programming problem, 
which is highly complex, due to the above-mentioned interference terms
causing the flow rates to have strong interdependencies, and due to the 
large number of discrete optimization variables involved. To the best of the
authors' knowledge, this formulation of routing problem (which accounts for
physical-layer interference) has not been studied previously in the literature.

\subsection{Motivation for Reinforcement Learning}

Optimizing routing and spectrum access in even just a single data flow is
already nontrivial: it involves sequential decision makings in a highly
variable environment, so it can be modeled as a Markov decision process.  
But, because each agent can only observe local information, and further due to
the lack of central infrastructure for control and communication in ad-hoc
networks, the optimization across the multiple interacting flows needs to be
conducted in a distributed manner, the problem of distributed routing and
spectrum access over the multiple data flows belongs to the class of
\emph{multi-player partially observable Markov decision process} (MP-POMDP).
This is a challenging problem, and it motivates the use of distributed
multi-agent reinforcement learning. 

Moreover, as the wireless environment is characterized by physical layer 
attributes modeled by continuous quantities, function approximators are
required for implementing the reinforcement learning agent. This motivates 
us to utilize deep reinforcement learning techniques. 

The novel approach taken in this paper is to
let a single reinforcement learning agent optimize one flow at a time by moving
along the \emph{frontier node} of the flow. Then, we optimize the next flow, and do
so successively, in effect letting flows compete with each other. As shown later in
the paper, this would already lead to effective solutions to the problem
(\ref{equ:problem}) for common network utilities such as the sum rate or the minimum rate.

\section{Flow based Deep Reinforcement Learning for Routing and Spectrum Access}\label{sec:method}

This section describes the proposed novel reinforcement learning approach for 
the multi-flow joint routing and spectrum-access problem.

\subsection{Single Agent for Multiple Nodes in a Flow}\label{sec:assoc}

\begin{figure}
    \centering
    \ifOneColumn
        \includegraphics[width=0.55\textwidth]{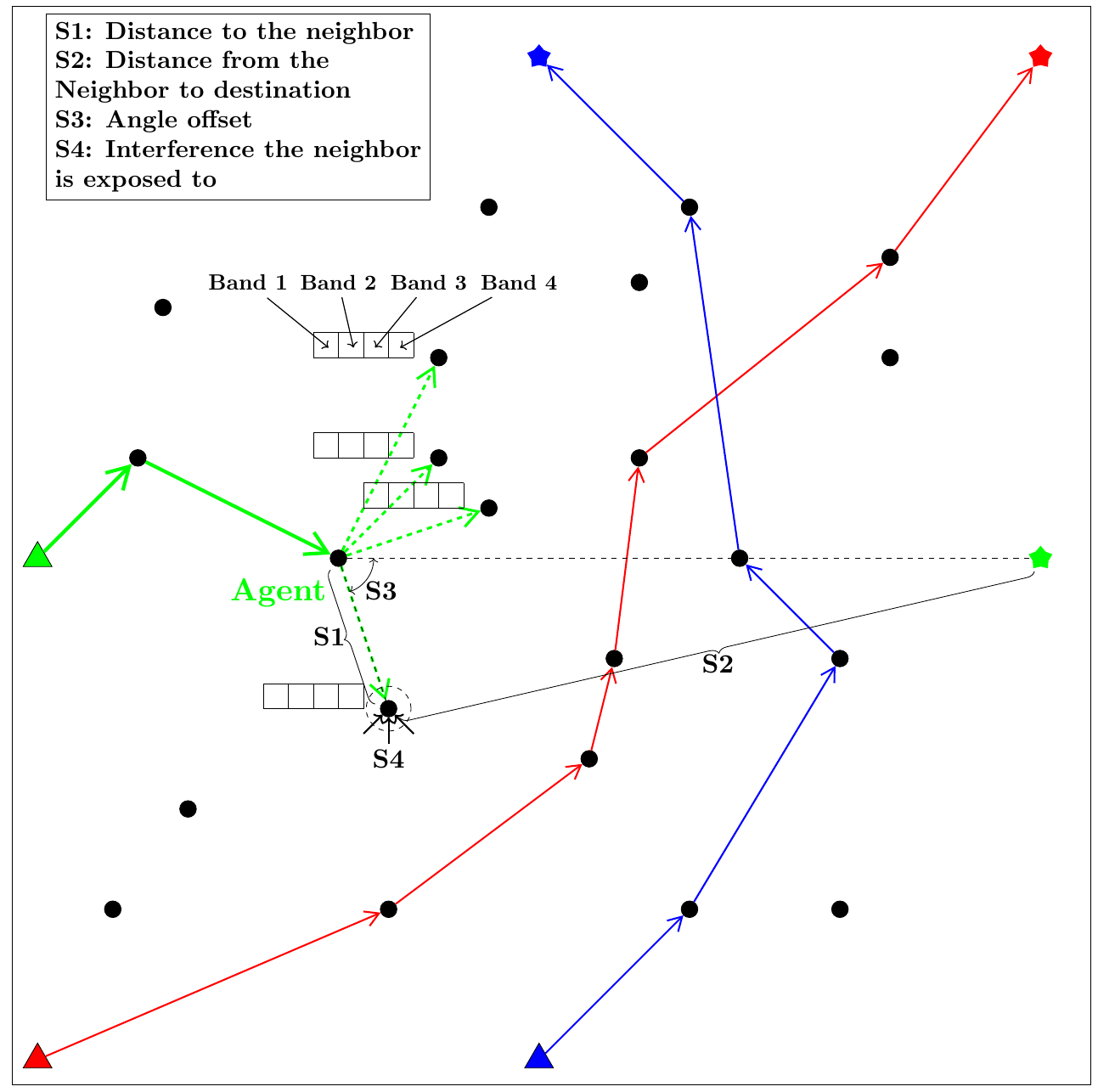}
    \else
        \includegraphics[width=0.4\textwidth]{Figures/Figure_Routing_Agent.pdf}
    \fi
    \caption{A reinforcement learning agent at the frontier node of the flow determines the best next-hop and the best frequency band based on state values S1 to S4 for each of $c$ neighbors in each of $B$ frequency bands.}
    \label{fig:agent}
\end{figure}

A key innovation of this paper as compared to prior work is to associate a
single reinforcement learning agent to all the nodes in each flow. 
The idea is that as a route is being established on a hop-by-hop basis, 
the agent moves along with the frontier node of the partially
established route to decide the best next hop as well as the best frequency 
band to reach the next node from the frontier node. Compared to most prior
approaches that train a different agent for each node in the network to perform
either routing or spectrum access, this \emph{agent-to-flow} association has
the key advantage that it significantly reduces the number of parameters that
need to be trained, because a single agent now works for all nodes in the flow.
This improves scalability, and further also improves generalization ability
because the same agent can be used across multiple data flows and even
across distinct ad-hoc networks. 
Fig.~\ref{fig:agent} provides a visualization of the proposed
agent-to-flow association for establishing the next hop at a frontier node of a flow. 

When optimizing multiple flows, we adopt a \emph{sequential} ordering for establishing routes and spectrum allocation: i.e., we let one flow establish its entire route and frequency band selections before the next flow, and do so over multiple rounds to allow the interference pattern to be fully established. 

\subsection{Single Agent for Multiple Frequency Bands}\label{sec:spectrum_access}

We aim to develop a reinforcement learning agent to simultaneously perform
routing and spectrum access for all the hops in each flow. One
natural design for the agent is to aggregate features from all frequency bands,
and to decide the optimal \emph{next node and frequency band} combination.
This, however, significantly increases the resulting state and action spaces
for the agent.  
Furthermore, this design does not allow the agent to be generalized to
scenarios with different number of frequency bands.

\begin{figure*}
    \centering
    \includegraphics[width=0.7\textwidth]{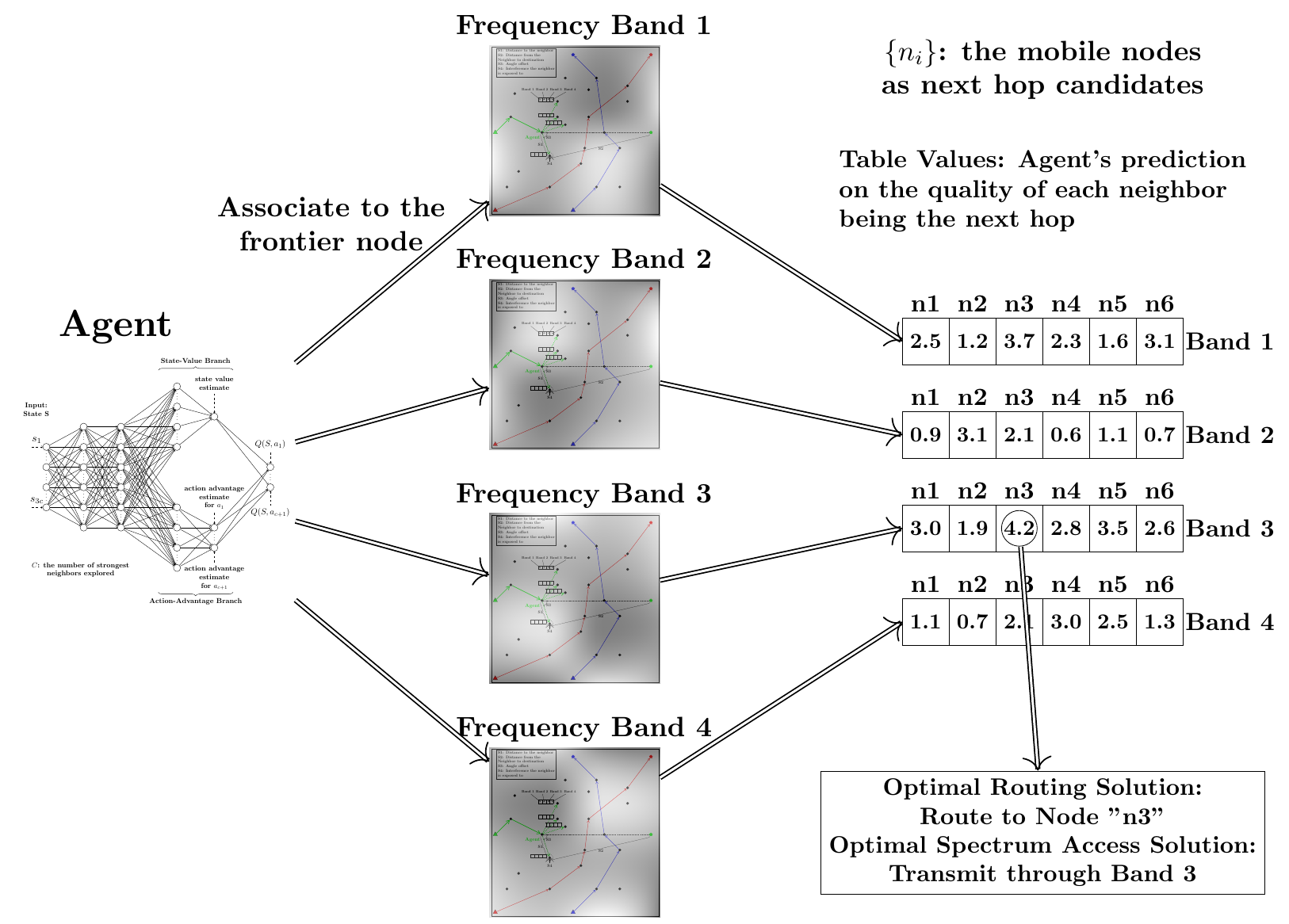}
    \caption{The same reinforcement learning agent is reused across all the frequency bands. The optimal next-hop node and frequency combination is determined from the outputs of the agent.}
    \label{fig:routemultibands}
\end{figure*}

To reduce the state space and the action space and to achieve higher model parameter
efficiency, we propose an approach of designing an agent for a single frequency
band, then replicating it across $B$ bands. 
The agent evaluates the suitability of the next node for all frequency bands, one band at a time, 
then selects the best node and frequency band combination. 
This process is illustrated in Fig.~\ref{fig:routemultibands}. In essence, the
agent achieves equivalent state space and action space over all the frequency
bands, with parameter cost of just one band. 

\subsection{Action Space and State Space}\label{sec:stateactionspace}

We now define the action space and the state space for the
reinforcement learning agent for one frequency band. The agent's action corresponds to selecting a next-hop node to transmit.
As this paper adopts a physical-layer network model in which the connections
are not predetermined, we consider a fixed number of $c$ available neighbors with
strongest channels as candidates. In addition, to ensure sufficient exploration
capability, we also add one \emph{reprobe} action, which means that if the
agent decides none of the $c$ strongest neighbors is a suitable next hop, it would proceed to probe the next $c$ strongest
neighbors, until a suitable next hop is found. Thus, over a single frequency band, 
the agent's action space $\mathcal{A}$ consists of $c+1$ actions: 
\begin{enumerate}[label=\textbf{A.\arabic*}] 
\item Total of $c$ actions: one for selecting each of the $c$ strongest neighbors.
\item One additional action for reprobing beyond the $c$ strongest neighbors.
\end{enumerate}

Note that if the agent chooses to reprobe, the agent stays at the current frontier node in the current time. 
At the next time step, the agent will explore the next $c$ closest neighbors and collect the state information corresponding to these new set of neighbors. 

The agent's state space $\mathcal{S}$ over one frequency band should summarize
crucial factors about each of the $c$ strongest neighbors in that frequency
band as the agent moves along the frontier node. We adopt a pathloss based network
model in which the SINR is a function of the geographical locations of the
neighbors relative to the frontier node and to the destination, as well as the
interference each neighbor is subjected to. To this end, the agent gathers the
following information from each neighbor to form its state:
\begin{enumerate}[label=\textbf{S.\arabic*}] 
\item Distance between the neighbor and the frontier node.\label{state:1}
\item Distance between the neighbor and the destination.  \label{state:2}
\item Angle difference between the direction from the frontier node to the neighbor and the direction from the frontier node to the destination. \label{state:3}
\item Total interference the neighbor is exposed to in the current frequency band.\label{state:4}
\end{enumerate}
Here, \ref{state:2} and \ref{state:3} indicate the amount of progress toward the 
destination node that can be made if the neighbor is chosen as the next hop, while
\ref{state:1} and \ref{state:4} together allow an estimation of the SINR to the
neighbor in the current frequency band. In total, the state over one frequency band is a $4c$-component vector. Fig.~\ref{fig:agent} shows the four state features for each of the $c$ closest neighbors.

Before we complete the picture by defining the agent's reward, we next provide a brief overview of reinforcement learning, and illustrate why the standard \emph{Q-Learning} needs to be modified in order to tackle the routing problem.

\subsection{Conventional Q-Learning}
\label{sec:qlearning}

Q-learning \cite{qlearning} is an off-policy, \emph{value-based} reinforcement
learning algorithm for \emph{model-free} control. Q-learning is suited for 
routing and spectrum access for wireless ad-hoc networks, because: 
\begin{itemize}
\item Being a value-based algorithm, Q-learning estimates a value for each
state-action pair. In the routing problem, each action corresponds to a
combination of next hop node and frequency band selection. These actions can be
quantitatively evaluated and compared based on the overall network objective
(such as the bottleneck rate of the flow). The value-based Q-learning algorithm
then allows these quantities to be mapped and learned depending on the
state-action values.  
\item Being a model-free algorithm, Q-learning allows us to bypass explicit
modeling and directly determine the optimal policy. This is useful, because in our setting, the state of the network is only partially observed by the agent, which makes the state transitions of the network difficult to model. 
\end{itemize}
The conventional Q-learning algorithm is based on the following definition of the 
\emph{action value} denoted as $Q$:
\begin{align}
     Q_\pi(s_t, a_t) =\:&\mathbb{E}_{\pi}[r_t+\gamma r_{t+1}+\gamma^2 r_{t+2}+\dots \ | \ s_t,a_t] \label{equ:futurerewardmontecarlo} \\
     =\:& \mathbb{E}_{\pi}[r_t + \gamma Q_\pi(s_{t+1},a_{t+1}) \ | \ s_t,a_t] \label{equ:futurerewardbootstrap}
\end{align}
where $r_t$ is the reward at time $t$ and $\gamma$ is a discount factor.
Here, $Q_\pi(s_t, a_t)$ is the expected \emph{future cumulative reward} after the agent executes the action $a_t$ from the state $s_t$, with the expectation taken under certain policy $\pi$. 
Given an estimate of $Q_\pi(s_t, a_t)$, one can improve the policy by acting greedily with respect to $Q_\pi(s_t,a_t)$, i.e., 
\begin{align}\label{equ:greedyimprov}
    \pi'(s_t) = \text{argmax}_a Q_\pi(s_t, a).
\end{align}
We can then re-estimate the values of $Q_\pi'(s_t, a_t)$ through either 
(\ref{equ:futurerewardmontecarlo}) or (\ref{equ:futurerewardbootstrap}), 
and continue to improve the policy using (\ref{equ:greedyimprov}). 
When the process converges, we reach an optimal policy $\pi^*$ 
(and its corresponding value functions $Q_{\pi^*}(s_t, a_t)$). 
Such a process is referred to as the \emph{policy iteration}, and 
the optimal policy takes on the following self-referential relation: 
\begin{align}
    \pi^*(s_t) = \text{argmax}_a Q_{\pi^*}(s_t, a).
\end{align}

When estimating $Q$ (which is commonly referred to as policy evaluation), there are two common approaches, each with its own benefits and shortcomings:
\begin{enumerate}
    \item Monte-Carlo Estimation, corresponding to (\ref{equ:futurerewardmontecarlo}), which provides an unbiased estimation, but has high variance;
    \item Temporal Difference Estimation, corresponding to (\ref{equ:futurerewardbootstrap}), which has low variance, but is biased due to its self-referential nature.
\end{enumerate}

The standard Q-learning algorithm performs the policy evaluation with temporal difference estimation, and updates the $Q$ estimation based on the Bellman optimality equation \cite{bellman} as follows:
\ifOneColumn
    \begin{align}
        Q^{\rm new}_{\pi}(s_t, a_t) = Q^{\rm old}_{\pi}(s_t, a_t) +
    	\alpha(r_t+\gamma \max_{a} Q^{\rm old}_{\pi}(s_{t+1}, a)-Q^{\rm old}_{\pi}(s_t, a_t)), \label{equ:updateactionvalue}
    \end{align}
\else
    \begin{multline}
        Q^{\rm new}_{\pi}(s_t, a_t) = Q^{\rm old}_{\pi}(s_t, a_t) + \\
    	\alpha(r_t+\gamma \max_{a} Q^{\rm old}_{\pi}(s_{t+1}, a)-Q^{\rm old}_{\pi}(s_t, a_t)), \label{equ:updateactionvalue}
    \end{multline}
\fi
where $\alpha$ is the step size. 
To accelerate the algorithm, Q-learning only performs (\ref{equ:updateactionvalue}) once before the policy improvement in each round of the policy iteration. 
(One step of the policy evaluation in a policy iteration process is also known as the \emph{value iteration} process.)

There are, however, two major issues when applying the standard Q-learning
algorithm to the routing and spectrum access problem for the wireless ad-hoc
network. First, the conventional Q-learning assumes a reward definition $r_t$ as 
the \emph{immediate} reward that the agent receives after executing action $a_t$
in state $s_t$. These immediate rewards form the cumulative reward $Q(s_t, a_t)$,
which is the objective that the agent aims to optimize. However, in the routing problem, the objective 
(\ref{equ:problem}) is determined by the bottleneck rates in the flows. Decomposing 
the bottleneck rate into a set of immediate rewards is not straightforward.

Secondly, 
even if the instantaneous rewards are redefined to account for the optimization
of the bottleneck rate, the standard training process for Q-learning, which is
based on temporal difference estimation, is known to suffer from high biases.
In our experience for training the reinforcement learning agent for routing
with the instantaneous rewards redefined to account for the bottleneck, the
learned agent's state-action value estimations are noticeably off with large
positive biases throughout the state space (even when using the Double-DQN
method \cite{doubledqn}), rendering the agent incapable of producing high
quality solutions for routing or spectrum access. 

\subsection{Novel Reward and Action Value Estimation}\label{sec:reward}

To apply Q-learning to the routing problem, we need first to incorporate the bottleneck rate into the reward
definition. The most intuitive way to do so is to set the reward to be all zero 
along the route until the last hop, which takes a reward value of (\ref{equ:bottleneckrate}), 
i.e., the minimum rate across all the hops in the route. However, this definition 
has the drawback that such a reward and the resulting Q-values for one hop could 
well be determined by an earlier hop, making them independent of the agent's 
choice of actions. Consequently, this makes it impossible for the agent to interpret 
the rewards during training.

To simultaneously address the problem of decomposing the bottleneck rate to
immediate rewards as well as the problem of biased estimation for temporal
difference state-action value estimation, this paper proposes a novel structure 
for rewarding the agent. Instead of identifying an instantaneous reward $r_t$
then computing $Q$ as the cumulative sum of $r_t$, we directly assign a future
cumulative reward for each state-action pair. Such a new definition of future
cumulative reward shares the same semantic meaning as the $Q$ function in
conventional Q-learning, but is not a result of cumulative instantaneous rewards.
The main advantage of this approach is that it makes possible to optimize network
objectives (such as the bottleneck rate in our case) that cannot be easily
decomposed as sum of instantaneous rewards.

Specifically, at any given node, we choose to use the \emph{bottleneck link
SINR from that node onwards in the route} as its future cumulative
reward.  Note there is a one-to-one relationship between rate and SINR.
More precisely, consider the data flow $f\in\mathcal{F}$ established in
(\ref{equ:dataflow}). During training, at frontier node $n^{(f)}_t, (0\leq
t\leq h)$, the agent observes the state $s_t$ and takes the action $a_t$ 
(leading to node $n^{(f)}_{t+1}$ using frequency band $b^{(f)}_t$). 
After the route is fully established, the state-action pairs along the route
are then assigned a future cumulative reward, which we denote as $\widetilde{Q}$, 
defined as follows:
\begin{align}\label{equ:qhat}
    \widetilde{Q}(s_t, a_t) = \min_{i=t\dots h}\text{SINR}_{(n^{(f)}_i, n^{(f)}_{i+1}, b^{(f)}_i)}.
\end{align}
We emphasize that $\widetilde{Q}$ plays the same role as $Q$ value in
Q-learning in that it incorporates future information (and no past information)
and is used as the metric for determining the optimal actions.

In the works on classical Q-learning \cite{qlearning} and deep Q-learning 
\cite{DQN,dueling,doubledqn}, the target for $Q(s_t,a_t)$ is computed based on
the temporal difference estimation with bootstrapping, which can cause high
biases as already mentioned. Instead, the $\widetilde{Q}$ defined in
(\ref{equ:qhat}) can be computed directly based on Monte-Carlo estimation,
which is unbiased. This is another crucial advantage of the proposed approach.

We note that Q-Learning is not originally designed to work with Monte-Carlo 
estimation. The above unconventional algorithm design, when used in conjunction with 
\emph{experience replay} for deep reinforcement learning, does give rise to one slight technical difficulty: each time the $\widetilde{Q}$ is
updated, it corresponds to the policy evaluation of a slightly older version
of policy (from which the samples are collected for Monte-Carlo estimation). 
In the next section, we provide a variation to the experience
replay technique to address this difficulty.

\subsection{Deep Q-Learning Based Routing and Spectrum Access}
\label{sec:dqnmodel}

Since the states are continuous variables, the agent needs to be able to
generalize over unseen portions of the state space $\mathcal{S}$. To this end,
we utilize deep reinforcement learning, specifically deep Q-learning
\cite{DQN}, in which a neural network, namely \emph{DQN}, is trained to predict
$Q$ values (in our case, the $\widetilde{Q}$ values) given the state-action
inputs. 

\begin{figure}
    \centering
    \ifOneColumn
        \includegraphics[width=0.55\textwidth]{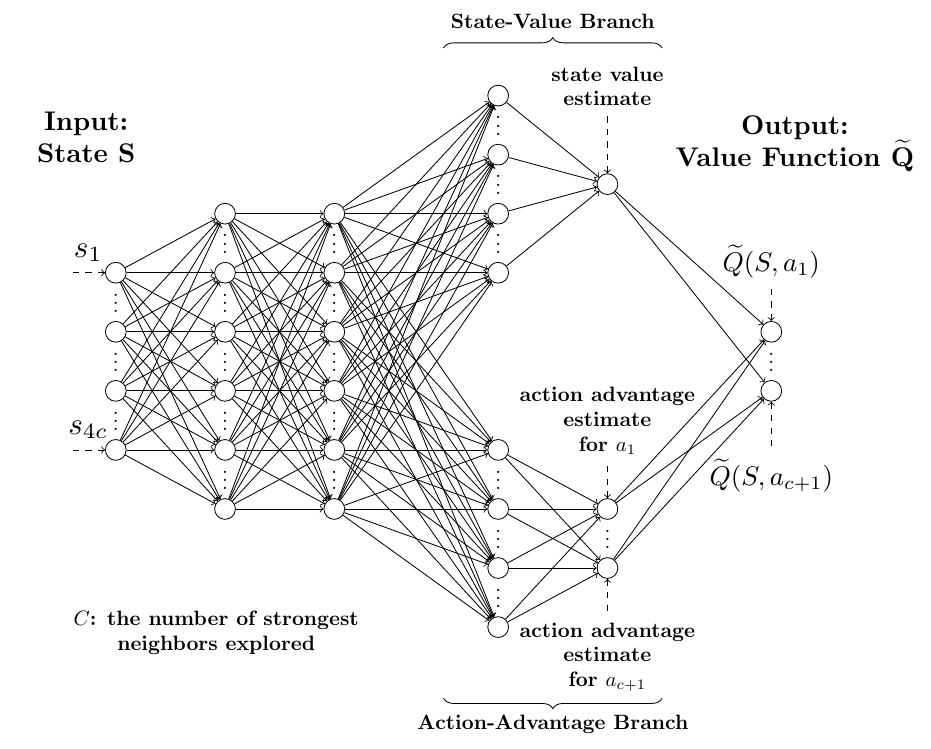}
    \else
        \includegraphics[width=0.48\textwidth]{Figures/Figure_Dueling_DQN_Agent.pdf}
    \fi
    \caption{Dueling-DQN Structure}
    \label{fig:DDQN}
\end{figure}

We format each state input for a single frequency band as a vector of length
$4c$ (four features for each of the $c$ strongest neighbors). These input
vectors are processed with fully connected layers with non-linearities.  We
adopt the state-of-the-art \emph{dueling-DQN} network architecture
\cite{dueling} for our agent, as illustrated in Fig.~\ref{fig:DDQN}.  The
dueling DQN consists of two separate estimators: one for estimating the state
value and the other for estimating the action advantage. Such a
separation has been shown to be beneficial for the agent to learn the distinctions between actions. 

\begin{figure*}
    \centering
    \includegraphics[width=0.75\textwidth]{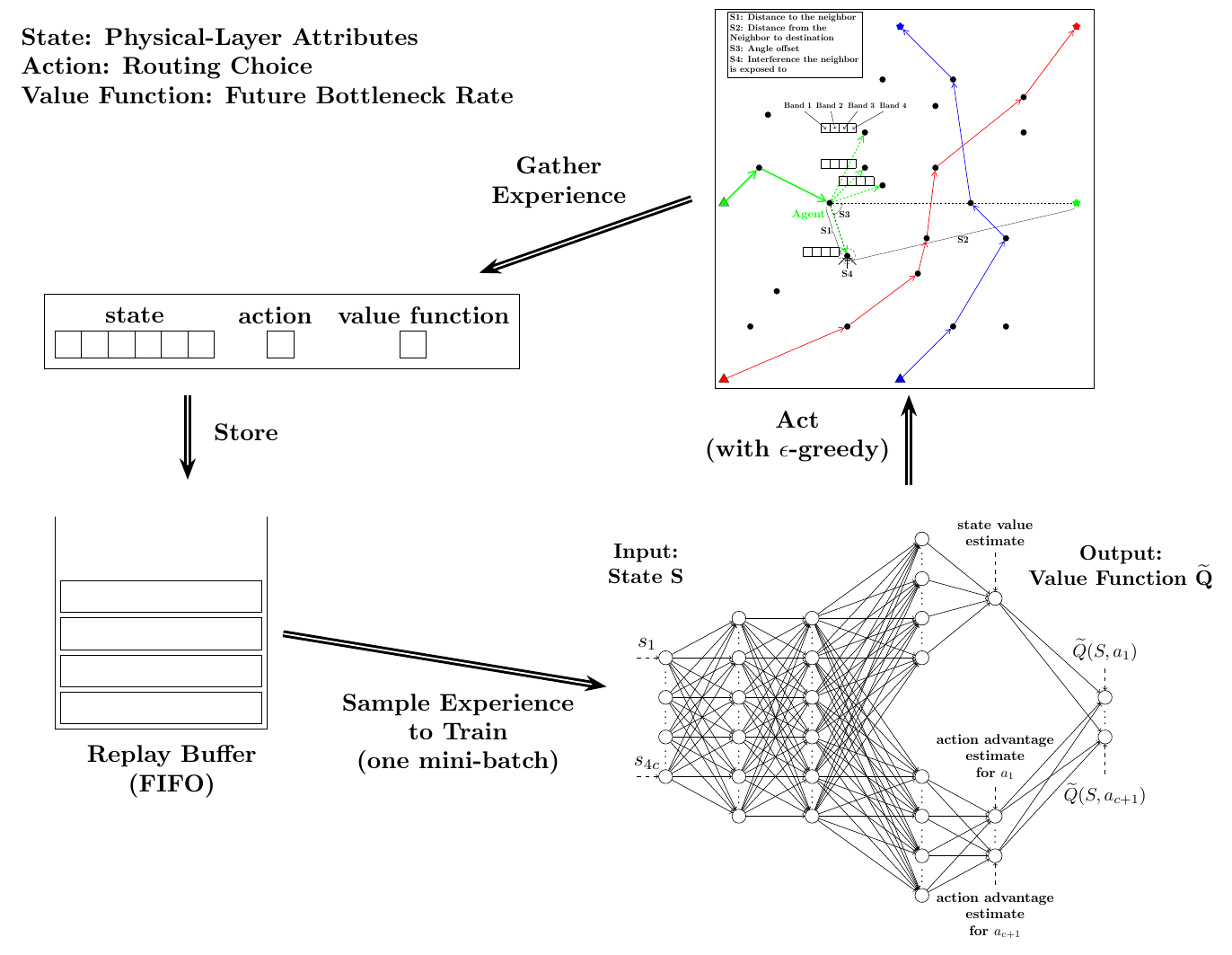}
    \caption{Training of Dueling-DQN with experience replay}
    \label{fig:replay}
\end{figure*}

The DQN is trained following the experience-replay technique as in \cite{DQN},
with uniform sampling from the replay buffer. The agent follows the
\emph{$\epsilon$-greedy} \cite{epsilongreedy} policy to gather experiences for
the replay buffer. More specifically, with probability $1-\epsilon$, the agent
acts greedily according to the agent's current estimation of $\widetilde{Q}$:
the action with the highest $\widetilde{Q}$ value across the $B$ frequency
bands is selected (as in Fig.~\ref{fig:routemultibands}).  With
probability $\epsilon$, the agent selects a random node as the next hop and a
random frequency band to reach that node.  If the selected node is not within
the $c$ strongest neighbors to the agent, we store a reprobing transaction. 
Together, this allows us to produce a series of agent transactions 
for routing from the source to the destination.
Once a route is fully established, we compute the rewards as defined by the
onward bottleneck SINR for each node in the route, and store the corresponding 
(state, action, reward) tuple in the replay buffer, which is used for subsequent DQN training.
Note that since the agent is trained to work in one frequency band at a time, 
the state is a $4c$-vector and the action is a value in $\{1, \dots, c+1\}$.
In case of greedy exploitation, we store the state vector for the frequency
band in which the highest $\widetilde{Q}$ is achieved; in case of random
exploration, we store the state vector for the randomly chosen frequency band. 
The same is true for the reprobe transactions.
The entire process of DQN training with experience replay is illustrated in Fig.~\ref{fig:replay}. 

We also propose a slight variation at the end of the training process to address the technical difficulty of using Monte-Carlo estimation for Q-learning. Normally, when training using the $\epsilon$-greedy policy, the agent always has a small (but non-zero) $\epsilon$ value, even at the end of the training process. To ensure that the policy evaluation is based on the most up-to-date policy, we set $\epsilon=0$ at the end of training, and let the agent further collect samples and train for an extended period of time, (which we refer to as the \emph{extended training}). With $\epsilon=0$, 
the agent's sampling policy reduces to the deterministic $\widetilde{Q}$ value-based greedy policy. Correspondingly, the Monte-Carlo estimation from these collected samples provides the policy evaluation on this most up-to-date deterministic policy, based on which the policy improvement can be computed. As the result, the policy iteration process from the conventional Q-learning is correctly recovered\footnote{Normally, Q-learning is regarded as a close variant to value iteration. However, if we regard each update step in Q-learning as an approximate policy evaluation, then we can still analyze Q-learning based on the policy iteration framework.}. 

To showcase the generalization ability of our method, we only train the agent
on one specific data flow during experience gathering.
Specifically, we first use a simple \emph{closest-to-destination
among the strongest neighbors} heuristic to produce routes of all but one data flow. We then
assign the agent to the last flow to gather
experiences and conduct training. As shown later in Section~\ref{sec:results},
such lightweight training strategy is already sufficient to produce an agent that
generalizes well across all data flows as well as across different ad-hoc networks. It does not need to be re-trained when the network topology changes, unlike in methods such as in \cite{chang}.

\subsection{Agent Policy Enhancements}\label{sec:enhance}
From domain knowledge, we propose two enhancements to the agent's policy:
\begin{itemize}
    \item If the agent chooses a neighbor that does not have the strongest channel to the frontier node, then all neighbors with stronger channels to the frontier node are excluded for future consideration for the next hop.
    \item If the destination node appears within the agent's exploration scope (i.e. one of the $c$ strongest neighbors), then the agent would not choose any neighbor which has a weaker channel to the frontier node than the destination node.
\end{itemize}
Both enhancements effectively prevent agents from taking non-essential back-and-forth hops. 

\subsection{Multi-Agent Distributed Optimization}\label{sec:rewardimp}

The agent learns and acts according to $\widetilde{Q}$ in (\ref{equ:qhat}) in
order to optimize the data rate of its own flow.  With each agent operating
under a \emph{selfish} objective, in general we do not expect the agents to
collectively converge to a global network-wide optimum. 
Fundamentally, achieving cooperation among distributed multi-agent is always a challenging problem. If we assign selfish rewards to the agents, then there is no guarantee that they would reach a global optimum; if instead we assign all agents a common global reward, we then encounter the \emph{credit mis-assignment} problem, as studied in \cite{optimalpayoff,localreward}. 

Furthermore, under the assumption that each agent only has local observations,
it is helpful for the agents to coordinate their actions. However, it is
challenging to engineer a messaging scheme that induces cooperation, while
incurring only limited communication overhead. Recognizing this difficulty, in
this paper, we let each agent learn independently, while being agnostic of the
other agents' actions or rewards, except through observing the surrounding
interference (essentially regarding the other agents as a part of the environment).
Despite its simplicity, a strategy like this can already be effective in many
situations \cite{idagent}.

We observe that in a wireless ad-hoc network, where the major performance
limiting factor is the mutual interference, routing optimization for one
flow in the presence of other flows would already tend to produce an interference avoidance behavior, which 
benefits other data flows as well. Therefore, designing the agents to act
selfishly in multiple rounds already allows every agent to adjust according to
each other's routes, thus achieving implicit cooperation to certain degree. The
effectiveness of this approach is illustrated in the simulation results in
Section~\ref{sec:results}.

\subsection{Fairness Among Data Flows}\label{sec:fairness}
In many applications, fairness among the data flows is an important design objective. 
We empirically observe that if routing and spectrum access are established 
sequentially across the data flows in multiple rounds, the data flows established 
relatively later tend to have better bottleneck rates. The reason is that 
the later flows can adjust according to the interference levels and utilize 
spectrum more efficiently. Therefore, to promote fairness among the data flows, 
we propose the following: At the end of each round, we compute the bottleneck 
rates for all data flows, then optimize routing and spectrum for the data flows in 
the next round in the decreasing order of the bottleneck rates. Flows with weak bottleneck rates
in the previous round obtain the advantage of being able adjust their routes and 
frequency bands later in the next round, and as a result achieve better data rates. 
Note that such sequential routing strategy requires some level of synchronization 
among the flows.


\section{Simulation Results}\label{sec:results}

\subsection{Ad-hoc Network Setup}\label{sec:origin_setting}
We consider wireless ad-hoc networks in a 1000m$\times$1000m region with $F=3$ data flows. The data flows are to be routed with $B=8$ available frequency bands, with each band having 5MHz bandwidth. The node density profile is specified over nine equally divided sub-regions, with $(6, 8, 7, 6, 5, 10, 8, 9, 6)$ nodes located randomly within each sub-region. We consider the short-range outdoor model ITU-1411 with a distance-dependent path-loss to model all wireless channels, over all frequency bands at 2.4GHz carrier frequency. All antennas have a height of 1.5m and 2.5dBi antenna gain. We assume a transmit power level of 30dBm for all nodes; the background noise level is at -130dBm/Hz.

We randomly generate 290,000 ad-hoc network layouts to train our agent. Among these, 20,000 layouts are used for random exploration based routing to generate the agent's initial experience, 250,000 layouts are used for the $\epsilon$-greedy policy based routing to train the agent, and the remaining 20,000 layouts are used for extended training with $\epsilon$ set to 0. 
For testing, we randomly generate 500 new ad-hoc network layouts according to the same distribution.

\subsection{Agent and Reward Specification}\label{sec:trainingspecs}

We use $c=10$ as the number of the strongest neighbors the agent explores each time. Correspondingly, the inputs to the dueling-DQN are 40-component vectors. The state inputs are processed by sequential fully-connected layers, organized into a state-value prediction branch and an action-advantage prediction branch, as in \cite{dueling}. The specifications for the neural network are summarized in Table~\ref{tab:neuralnetspecs}.  

\begin{table}[t]
\caption{Design Parameters for the DDQN Agent}
\centering
\begin{tabular}{|l|l|c|}
\hline
Parameters & \multicolumn{2}{c|} {Number of Neurons}  \\
\hline
\multirow{2}*{\shortstack[l]{Initial Main-Branch\\ fully-connected layers}}
& 1st & 150 \\ \cline{2-3}
& 2nd & 150 \\ 
\hline
\multirow{2}*{\shortstack[l]{State-Value \\ fully-connected layers}} 
& 1st & 100 \\ \cline{2-3}
& 2nd & 1 (1 state value) \\ 
\hline
\multirow{2}*{\shortstack[l]{Action-Advantage \\ fully-connected layers}} 
& 1st & 100 \\ \cline{2-3}
& 2nd & 11 (11 actions) \\ 
\hline
\end{tabular}
\label{tab:neuralnetspecs}
\end{table}

This paper uses the bottleneck SINR onwards in the route as the
future cumulative reward $\widetilde{Q}$ assigned to each state-action pair.
For practical training, we express SINR in dB scale in order to have a more
suitable range of values. Further, we add a constant bias to ensure that the
reward is almost always positive. This is so that this reward definition can be 
extended to the case in which a discount factor related to delay can later be added. 
Specifically, in actual training, we define 
the $\widetilde{Q}$ function in term of SINR as follows: 
\begin{align}\label{equ:normalizesinr}
    \widetilde{Q}(s_t, a_t) = \min_{i=t\dots h}
	10\log_{10}\left(\text{SINR}_{(n^{(f)}_i, n^{(f)}_{i+1}, b^{(f)}_i)}\right) + \mathrm{bias}.
\end{align}
This makes the reward much easier to learn.

\subsection{Sum-Rate and Min-Rate Performance}\label{sec:sumrate}

We present the sum-rate and min-rate testing results for deep reinforcement learning agent (i.e., the dueling-DQN, or DDQN) as compared to a comprehensive list of benchmarks. These benchmarks are greedy in nature, since to our knowledge there currently do not exist efficient globally optimal algorithms for ad-hoc network routing based on physical-layer attributes. The benchmarks are as following:
\begin{itemize}
    \item \textit{Strongest neighbor: } Select the neighbor with the strongest wireless channel from the frontier node. 
    \item \textit{Best direction among neighbors:} Select the neighbor with the best next hop direction (i.e. the smallest angle offset between the agent-to-neighbor direction and the agent-to-terminal direction). 
    \item \textit{Closest to destination among neighbors:} Select the neighbor closest to the destination node. If all neighbors are further away from the destination than the agent, the agent will reprobe. 
    \item \textit{Least interfered among neighbors: } Select the neighbor exposed to the lowest amount of interference. 
    \item \textit{Largest data rate among neighbors: } Select the neighbor with the highest link capacity. 
    \item \textit{Destination directly: } Directly route from the source to the destination node.
\end{itemize}
These benchmarks incorporate a comprehensive set of important criteria, including but not limited to all the state features \ref{state:1}-\ref{state:4}. 

\begin{figure*}
    \centering
        \includegraphics[width=0.7\textwidth]{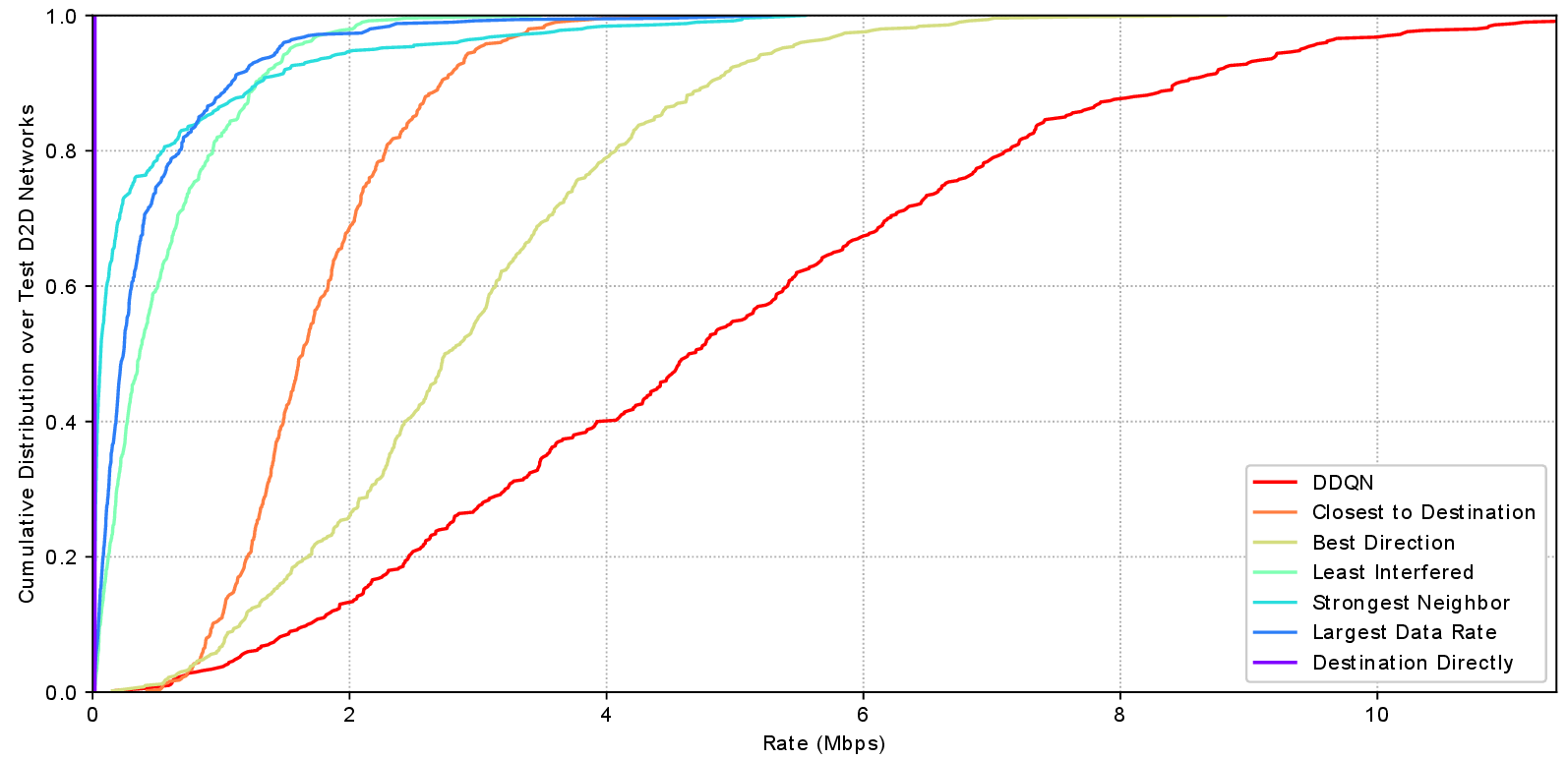}
	\caption{Cumulative distribution function of sum rate over 3 flows in 1000m$\times$1000m networks.}
    \label{fig:sumrate}
\end{figure*}

We follow the multi-round sequential routing for all the methods. 
In simulations, two rounds appear to be sufficient already. For the DDQN agent,
spectrum access is solved as in Section~\ref{sec:spectrum_access}. For all the
benchmarks, we select the best frequency bands based on the state feature \ref{state:4}. 
The sum-rate and min-rate results are summarized in
Table~\ref{tab:sumrate}. The CDF curves of sum-rate results are presented in
Fig.~\ref{fig:sumrate}. As shown, our proposed method achieves significantly
better performance than all benchmarks in both sum-rate and min-rate
performances. We emphasize that this performance is achieved by re-using a
single agent across all data flows, across all frequency bands, and over all
testing network layouts.

\begin{table}[t]
	\caption{Average Sum-Rate and Min-Rate Performances (Mbps)}
\centering
\ifOneColumn
    \begin{tabular}{|c|c|c|}
    \hline
    Methods & Sum Rate & Min Rate \\[2pt]
    \hline
    DDQN agent & 4.90 & 0.619 \\
    \hline
    Best Direction & 2.92 & 0.352 \\ 
    \hline
    Closest to Destination & 1.76 & 0.238 \\
    \hline
    Least Interfered & 0.55 & 0.040 \\
    \hline
    Largest Data Rate & 0.44 & 0.023 \\ 
    \hline
    Strongest Neighbor & 0.43 & 0.007 \\
    \hline
    Destination Directly & 0.02 & 0.003 \\
    \hline
    \end{tabular}
\else
    \begin{tabular}{ccc}
    \hline
    \topstrut Methods & Sum Rate & Min Rate \\[2pt]
    \hline
    \topstrut DDQN agent & 4.90 & 0.619 \\
    \topstrut Closest to Destination & 1.76 & 0.238 \\
    \topstrut Best Direction & 2.92 & 0.352 \\
    \topstrut Largest Data Rate & 0.44 & 0.023 \\ 
    \topstrut Strongest Neighbor & 0.43 & 0.007 \\
    \topstrut Least Interfered & 0.55 & 0.040 \\
    \topstrut Destination Directly & 0.02 & 0.003 \\
    \hline
    \end{tabular}
\fi
\label{tab:sumrate}
\end{table}

\begin{figure}
    \centering
    \includegraphics[width=0.48\textwidth]{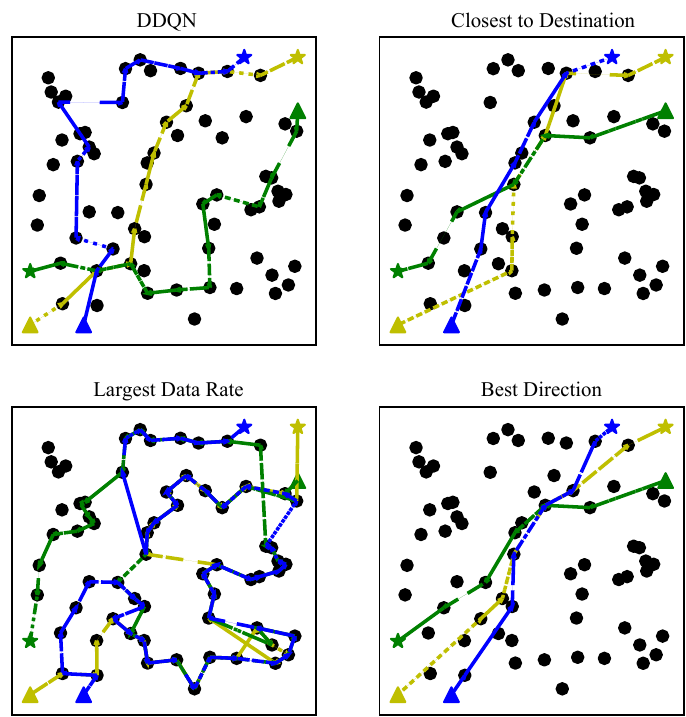}
    \caption{Routes achieved on three data flows. Data flows are differentiated by colors, while frequency bands are differentiated by line styles. The dimension of the area is 1000m$\times$1000m.}
    \label{fig:routes}
\end{figure}

To better understand how our method excels, we provide a visualization of routes formed by the DDQN agent together with selected benchmarks over a random ad-hoc network layout in Fig.~\ref{fig:routes}. As shown, our agent intelligently spreads out all data flows as well as switching frequency bands for transmission to mitigate mutual interference, while still maintaining strong and properly directed links to the destination to form the routes.

To further illustrate this point, in Table~\ref{tab:spectrumefficiency}, we
show the spectral efficiency in bps/Hz achieved when the number of available
frequency bands varies from 2 to 8. It is clear that the DDQN agent is able to
achieve higher spectrum efficiencies when more frequency bands are available.
These results show that the DDQN agent is able to intelligently utilize all
frequency resources and mitigate interference by allocating transmission powers
across different frequency bands.

\begin{table}[t]
    \centering
    \caption{Spectrum Efficiency Analysis for DDQN Agent}
    \begin{tabular}{ccc}
        \hline
        \topstrut $B$ & Sum Rate (bps/Hz) & Min Rate (bps/Hz) \\
        \hline
        \topstrut 2 & 0.066 & 0.006 \\
        \topstrut 4 & 0.312 & 0.032 \\
        \topstrut 8 & 0.984 & 0.124 \\
        \hline
    \end{tabular}
    \label{tab:spectrumefficiency}
\end{table}

\subsection{Generalization Performance}
\subsubsection{Generalization to Different Number of Data Flows}
We now investigate the generalizability of the proposed method when the number of data flows in the testing environment is different from the training set. We directly take the agent trained under the original setting and reuse it across all the data flows under the new test settings. The ad-hoc network layouts are generated under the same distribution of transmission nodes within the same area as in Section~\ref{sec:origin_setting}. However, the number of data flows to be routed and evaluated is set to be 2, 4, 5, and 6, which are different from the original setting of 3 data flows. The results are summarized in Table~\ref{tab:nflows}.
\begin{table}[t]
    \centering
    \caption{Average Sum-Rate Performances With Different Number of Flows (M{bps})}
    \begin{tabular}{ccccc}
    \hline
    \topstrut Number of Flows & 2 & 4 & 5 & 6 \\[2pt]
    \hline
    \topstrut DDQN agent & 5.19 & 5.26 & 5.28 & 4.73 \\
    \topstrut Best Direction & 2.54 & 3.78 & 4.29 & 4.30 \\ 
    \topstrut Closest to Destination & 1.21 & 2.98 & 3.55 & 3.60 \\
    \topstrut Least Interfered & 1.22 & 0.58 & 0.55 & 0.47 \\
    \topstrut Largest Data Rate & 0.46 & 0.50 & 0.44 & 0.39 \\ 
    \topstrut Strongest Neighbor & 0.31 & 0.31 & 0.26 & 0.24 \\
    \topstrut Destination Directly & 0.01 & 0.03 & 0.04 & 0.05 \\
    \hline
    \end{tabular}
    \label{tab:nflows}
\end{table}

Simulation results suggest that the proposed DDQN agent still maintains the best performance among all testing approaches, although its advantage gradually decreases with the higher number of flows. The decreasing performance advantage is likely due to the fact that with the higher number of flows, the interference distribution becomes considerably different from the training setting, so the experience learned from the training set may not be as applicable. Nonetheless, routing with the DDQN agent is still the most promising approach even under these generalized ad-hoc network settings.

\subsubsection{Generalization to Larger Networks}
\begin{figure*}
    \centering
        \includegraphics[width=0.8\textwidth]{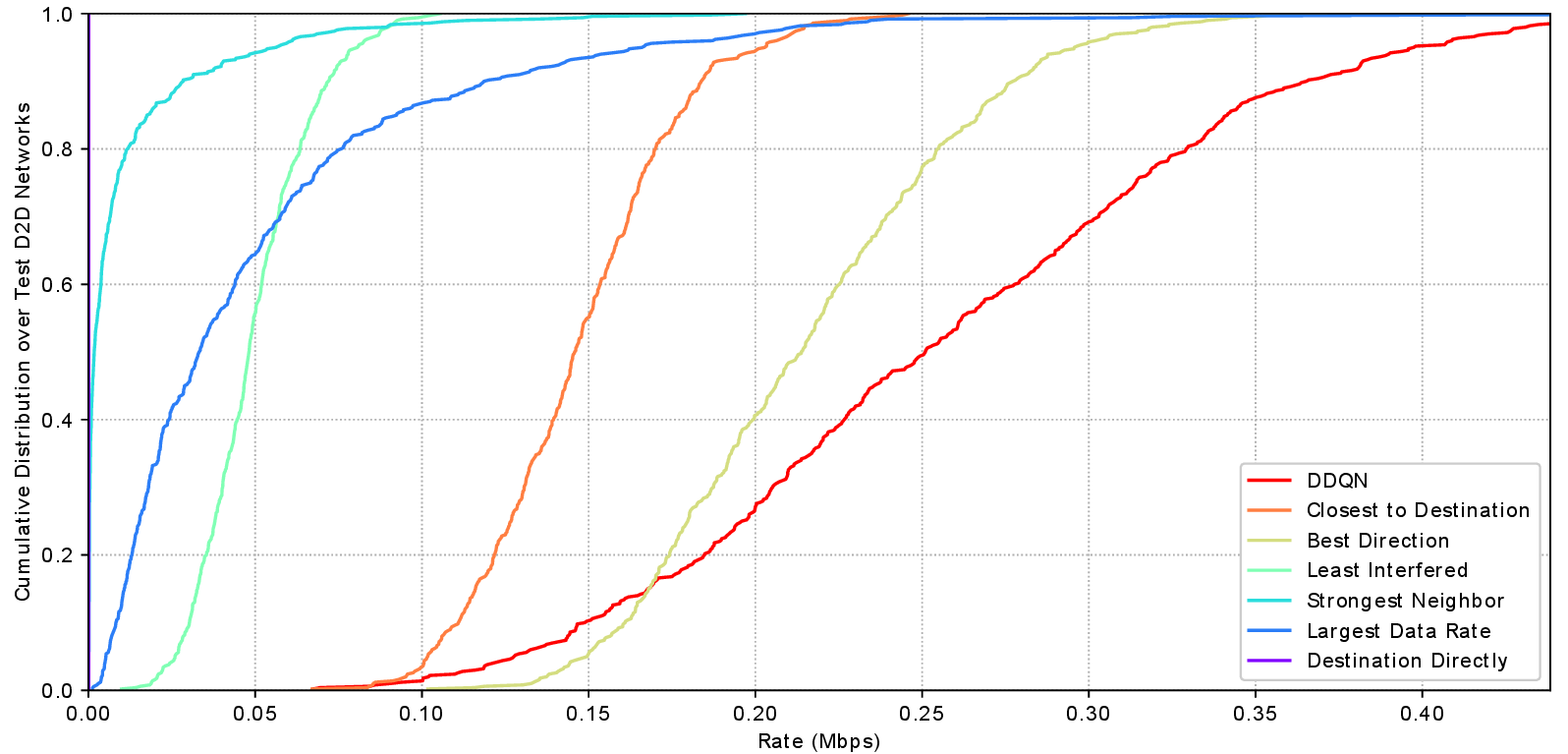}
	\caption{Cumulative distribution function of sum rate over 10 flows in 5000m$\times$5000m networks.}
    \label{fig:sumrate_large}
\end{figure*}

To test the generalizability of the proposed method on larger problem instances, we directly take the agent
trained under the original setting and reuse it for a much larger ad-hoc network of $F=10$ data flows in a 5000m$\times$5000m region, with $B=32$ available frequency bands. We place larger number of $(19, 16, 21, 18, 14, 24, 17, 20, 19)$ nodes over nine evenly divided sub-regions. The sum-rate results are shown in Fig.~\ref{fig:sumrate_large}, which shows that our approach still significantly outperforms the benchmarks. 

We notice that if we increase the number of relays nodes more aggressively, the
performance of the DDQN agent drops below the strongest benchmarks tested. The
potential reason is due to the fact that the DDQN agent has learned to prioritize
the relatively shorter links in our simulation setting. Using shorter links is
beneficial for obtaining higher direct-channel strengths, but it also leads to
more nodes being active, thus more interference in the entire wireless network.
As the DDQN agent is trained under the network settings as in
Section~\ref{sec:origin_setting}, it has learned the right balance between the
cost and benefit of using shorter links for achieving the optimal performance for
that setting.
But, when the network setting is drastically changed, the trade-off that the
agent had previously learned might no longer be correct. 

Nevertheless, the results presented in this subsection still illustrate that the proposed method has the capability to adapt to networks that are much larger in size, in the number of mobile nodes, and in the number of frequency bands, than the networks used for training. 

\subsection{Optimal Relaying Distance}

The heart of the routing and spectrum access problem is to
achieve the right balance between using nearby nodes as relays (which have 
strong channels) versus directly transmitting to farther-away nodes (which results in
fewer number of transmissions in the network hence lower overall interference). 
The DDQN agent learns this balance in a data driven fashion. 

For a network with a fixed node density, it is conceivable that a heuristics
can be designed based on identifying an optimal relay distance, so that it can
approach the performance of DDQN.  Indeed, Table~\ref{tab:linklength} shows the
performance of the ``closest-to-destination'' benchmark, but the set of
neighbors to explore is varied from 2 to 25. We see that the sum-rate and
min-rate performances are the best when 4 neighbors are explored, which
correspond to a medium relaying distance.  Although this benchmark still does
not outperform DDQN, its performance comes close.  The point is, however, that this optimal number of neighbors to explore is difficult to determine in
advance, while the proposed DDQN agent is able to identify the optimal number 
through training.

\begin{table}[t]
    \centering
    \caption{Closest-to-Destination Benchmark with Different Number of Neighbors Explored}
    \begin{tabular}{ccc}
    \ifOneColumn
        \hline
        \# of Neighbors Explored & Sum Rate (Mbps) & Min Rate (Mbps) \\
        \hline
        2 & 4.02 & 0.430\\
        4 & 4.56 & 0.517\\
        6 & 3.79 & 0.455\\
        8 & 2.67 & 0.311\\
        10 & 1.76 & 0.238 \\
        15 & 0.89 & 0.119 \\
        25 & 0.33 & 0.050\\
        \hline
    \else
        \hline
        \shortstack[c]{\topstrut\# of Neighbors \\ Explored} & \shortstack[c]{Sum Rate \\ (Mbps)} & \shortstack[c]{Min Rate \\ (Mbps)} \\
        \hline
        \topstrut 2 & 4.02 & 0.430\\
        \topstrut 4 & 4.56 & 0.517\\
        \topstrut 6 & 3.79 & 0.455\\
        \topstrut 8 & 2.67 & 0.311\\
        \topstrut 10 & 1.76 & 0.238 \\
        \topstrut 15 & 0.89 & 0.119 \\
        \topstrut 25 & 0.33 & 0.050\\
        \hline
    \fi
    \end{tabular}
    \label{tab:linklength}
\end{table}

In the same vein, it is also conceivable that an optimal spectrum access
heuristic can be designed. For example, we found empirically that if the agent
always chooses the
frequency band with the least amount of interference, it would generally work
very well. In fact, the DDQN agent chooses the frequency band with the least
interference about 70\% of the time. Nevertheless, the proposed Q-learning
based approach is still justified, because there is no easy way to come up with
these heuristics, especially when the choice of the frequency band needs to be
coupled with the choice of the next hop. Further, without DDQN, it is not
straightforward to evaluate the optimality of these heuristics.

\subsection{Fairness Among Flows}

\begin{figure}
    \centering
    \ifOneColumn
        \includegraphics[width=0.7\textwidth]{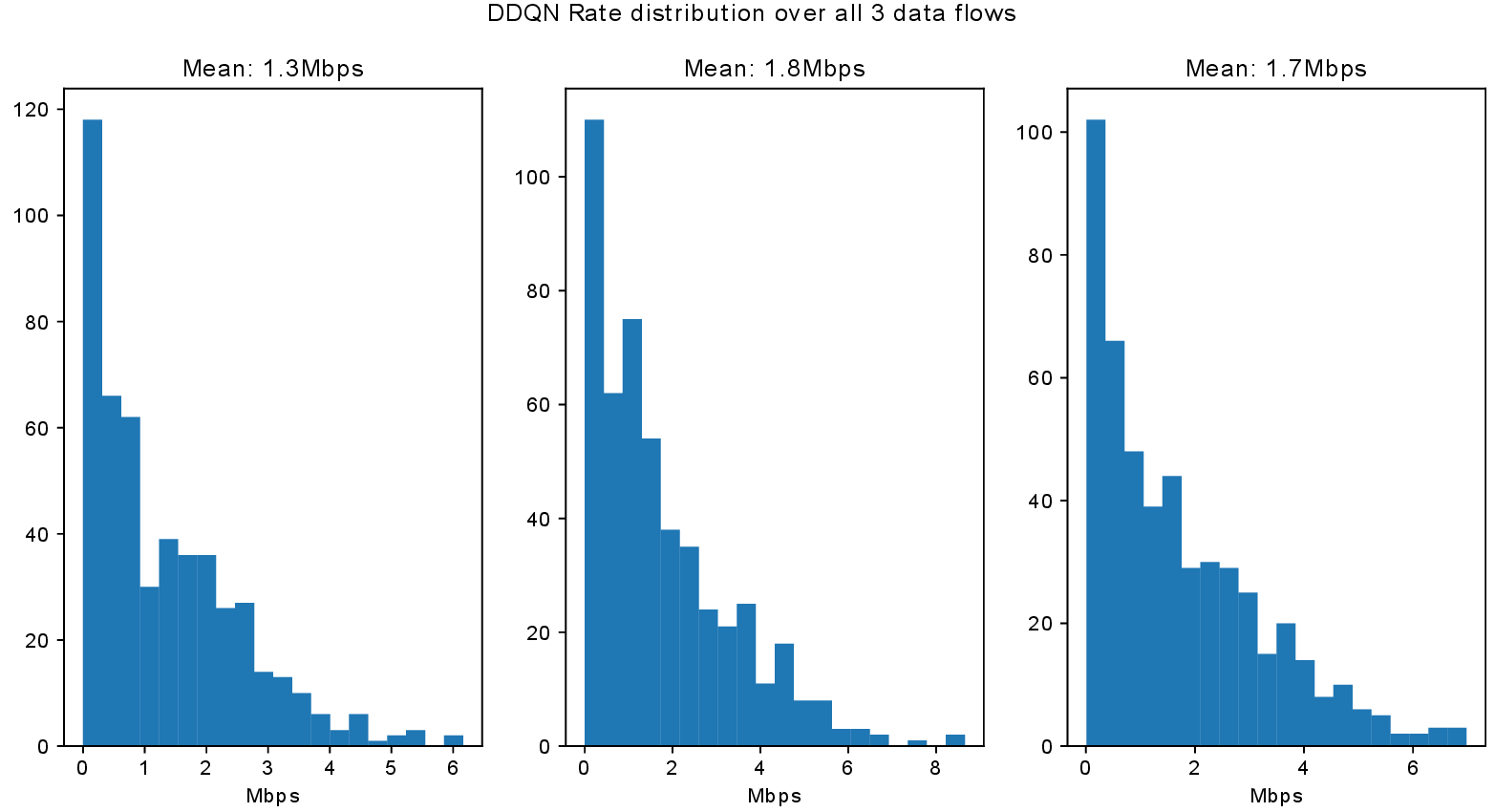}
    \else
        \includegraphics[width=0.47\textwidth]{Figures/rate_distribution}
    \fi
	\caption{Bottleneck rates over 3 data flows in 1000m$\times$1000m networks.}
    \label{fig:fairness}
\end{figure}

In the proposed DDQN, we encourage fairness between data flows by ordering
the optimization of the flows in the reverse order of the data rates
achieved in the previous round. To validate the achieved fairness using this approach, 
we provide numerical simulations of bottleneck rates achieved by each of the three flows
over multiple testing layouts drawn from the same setting. The results are
shown in Fig.~\ref{fig:fairness}. The distributions of the bottleneck rates are
very close for all three data flows, indicating that fairness is achieved over
the long run.

\section{Incorporating Delay in the Objective}\label{sec:delay}


Up to this point, we have exclusively focused on optimizing network utilities
based on the bottleneck rate of each flow. On the other hand, \emph{delay} is also an
important performance metric, which has been the focus of many prior routing work.
As an extension of the proposed approach, in this section, we show how delay
performance can be incorporated into the network objective, and propose training 
techniques for the reinforcement learning agent to achieve a tradeoff between the
bottleneck rate and time delay---two quantities which often conflict with each
other. 

Similar to most prior routing works, we quantify the transmission delay of
a given flow as the \emph{number of hops} from the source to the
destination. To incorporate the number of hops into the reward function 
of the DDQN agent, we propose to augment the reward function $\widetilde{Q}$ as
\ifOneColumn
    \begin{align}\label{equ:qhatdecay}
    	\widetilde{Q}(s_t, a_t) = \left\{\min_{i=t\dots h} 10\log_{10}\left(\text{SINR}_{(n^{(f)}_i, n^{(f)}_{i+1}, b^{(f)}_i)}\right) + \mathrm{bias} \right\} \cdot \lambda^{h-t}
    \end{align}
\else
    \begin{multline}\label{equ:qhatdecay}
    	\widetilde{Q}(s_t, a_t) = \\ \left\{\min_{i=t\dots h} 10\log_{10}\left(\text{SINR}_{(n^{(f)}_i, n^{(f)}_{i+1}, b^{(f)}_i)}\right) + \mathrm{bias} \right\} \cdot \lambda^{h-t}
    \end{multline}
\fi
where $h-t$ is the number of hops from node $n_{t+1}$ to the destination 
(i.e., remaining number of hops after executing action $a_t$); and $\lambda \in (0,1)$
is a penalizing factor. With $\lambda$ closer to
zero, the agent would prioritize more on minimizing time delays. 
Note that the purpose of the constant bias term is to ensure that the reward value is always positive so that it can be properly penalized by the penalty factor.

To explore such rate-delay
tradeoff, we train the DDQN agent with varying values of $\lambda$, each time from
scratch. Once the agent model is trained, we evaluate its performance on the testing ad-hoc
networks. The observed performance tradeoff between bottleneck rate and time
delay is shown in Table~\ref{tab:delaytradeoff}.  

\begin{table}[t]
	\caption{Tradeoff between bottleneck rate and time delay}
    \centering
    \ifOneColumn
        \begin{tabular}{|c|c|c|c|c|}
        \hline
        ${\lambda}$ & Sum Rate (Mbps) & Min Rate (Mbps) & \# of hops per data flow & \# of reprobes per data flow\\
        \hline
        ${1}$ & 4.90 & 0.619 & 13.2 & 0.01 \\
        \hline
        ${0.8}$ & 2.07 & 0.343 & 5.3 & 0.20 \\ 
        \hline
        ${0.6}$ & 0.31 & 0.050 & 2.9 & 5.00 \\ 
        \hline
        \end{tabular}
    \else
        \begin{tabular}{ccccc}
        \hline
        ${\lambda}$ & \shortstack[c]{Sum Rate \\ (Mbps)} & \shortstack[c]{Min Rate \\ (Mbps)} & \shortstack[c]{\topstrut \# hops \\ per flow} & \shortstack[c]{\topstrut \# reprobes \\ per flow}\\
        \hline
        \topstrut 1 & 4.90 & 0.619 & 13.2 & 0.01 \\
        \topstrut 0.8 & 2.07 & 0.343 & 5.3 & 0.20 \\ 
        \topstrut 0.6 & 0.31 & 0.050 & 2.9 & 5.00 \\ 
        \hline
        \end{tabular}
    \fi
\label{tab:delaytradeoff}
\end{table}

As shown in the results, the tradeoff is effectively achieved as $\lambda$ decreases from 1 to 0.
Specifically, with a low value of $\lambda$, the agent automatically learns to utilize the reprobe action
more aggressively to achieve fewer number of links per data flow. These results
suggest that with a properly modified reward definition and action space, the DDQN
agent can achieve a desired tradeoff between objectives that conflict with each other.
We note here that empirically when $\lambda < 0.8$, the DDQN agent reaches the destination on average faster than the \textit{Closest to destination among neighbors} benchmark.

\section{Benefit of Power Control}\label{sec:powercontrol}

This paper has so far assumed that each node always transmits at its fixed
power level. However, after determining the optimal routes and the set of
frequency bands, the bottleneck rates can be further improved if we allow
the transmission powers of the nodes to vary. 

Power control among interfering links is by itself already a highly challenging
optimization problem. The problem becomes even more complex if power control
is performed in combination with routing and optimal spectrum access. 
As this paper mainly focuses on routing, we will leave optimal power control
for future study. Instead, in this section, we simply apply
power control after the routes and frequency bands are established.

We propose the following straightforward and computationally efficient
heuristic for power control within each flow.  The heuristic is based on the
fact that only the bottleneck links affect the overall data rate of the flows. 
Therefore, for each of the remaining links, we can decrease its transmission power 
as long as its transmission rate does not drop below the bottleneck rate.
Decreasing these \emph{excessive} transmission power levels leads to less
interference, which helps improve the bottleneck rates. 

In details, we fix the transmission power for the bottleneck link, then reduce
the transmission powers for all other links to approximately align their
link rates to the bottleneck. Specifically, consider flow $f\in\mathcal{F}$ with the routing and
spectrum access solution as in (\ref{equ:dataflow}), with the 
transmission power levels as $\{p^{(f)}_{0}, p^{(f)}_{1} \dots p^{(f)}_{h}\}$. We then adjust
the transmission power for every node $n^{(f)}_{i}, 0\leq i\leq h$ as
\begin{align}\label{equ:powercontrol}
	p^{(f)}_{i,{\rm adjusted}} = \frac{\text{SINR}_\mathrm{bottleneck}^{(f)}}{\text{SINR}_{(n^{(f)}_{i},n^{(f)}_{i+1},b^{(f)}_{i})}}p^{(f)}_{i}
\end{align}
where the SINR values are computed as in (\ref{equ:SINR}), and $\text{SINR}_{\mathrm{bottleneck}}^{(f)}$ denotes the bottleneck SINR for $f$.

After adjusting the powers for one flow, the bottleneck SINRs for all other flows would increase due to the reduced interference. We recompute $\text{SINR}_{\mathrm{bottleneck}}^{(f)}$ for all the flows, and 
go through all the flows successively in one round, from the data flow with highest bottleneck SINR value to the lowest. 

While this method is heuristic, it is sufficient to illustrate the benefit of
power control. In Table~\ref{tab:powercontrol}, we provide a comparison of 
achieved rates before and after power control, with the routing and spectrum access
solutions already determined by the DDQN agent. As the results suggest, with even a
simple power control heuristic within each data flow, the bottleneck
rates can already be improved. These results show the efficacy of this
simple power control method even after the routes are already fixed, as well as the potential for more sophisticated rate optimization.  
\begin{table}[t]
	\caption{Power Control within Each Data Flow}
    \centering
    \begin{tabular}{ccc}
    \ifOneColumn
        \hline
        Methods & Sum Rate (Mbps) & Min Rate (Mbps) \\
        \hline
        No Power Control & 4.88 & 0.608 \\
        \hline
        With Power Control & 5.29 & 0.693  \\
        \hline
    \else
        \hline
        Methods & \shortstack[c]{\topstrut Sum Rate \\ (Mbps)} & \shortstack[c]{\topstrut Min Rate \\ (Mbps)} \\
        \hline
        \topstrut No Power Control & 4.88 & 0.608 \\
        \topstrut With Power Control & 5.29 & 0.693  \\
        \hline
    \fi
    \end{tabular}
\label{tab:powercontrol}
\end{table}

\section{Conclusion}\label{sec:conclusion}

This paper proposes a physical-layer based reinforcement learning approach to
the wireless ad-hoc network routing and spectrum access problem. By training a
universal agent along the different nodes in a flow and across different
frequency bands, and by using the physical environment as the input, together with 
a novel definition of reward function that accounts for the bottleneck rate
in the route, we arrive at a highly scalable algorithm for simultaneous routing
and spectrum access which can be adapted to the varying network layout
characteristics and generalized to larger networks. The numerical results
provide insight into the optimal relaying distance in an ad-hoc network
setting.  We further propose methods to provide fairness, as well as extensions
for incorporating delay into the optimization objective, and heuristic power
control methods that can further improve the overall performance of the network. 

The proposed method achieves scalability by recognizing that as long as the
physical environment around each node is relatively stationary, a single agent
can be trained to perform resource allocation tasks efficiently across the
entire network. Thus, for ad-hoc networks, just as shown in early work
\cite{cui_spatial} that the optimal scheduling strategy can be learned based on
the physical environment using a universal deep learning agent, this paper
shows that a universal reinforcement learning agent can learn the optimal
routing and spectrum allocation strategy based on physical inputs.  Together,
these results show the feasibility of using machine learning to solve
communication and networking problems that are otherwise difficult to tackle.

\bibliographystyle{IEEEtran}
\bibliography{IEEEabrv,citations}

\begin{thebibliography}{10}
\providecommand{\url}[1]{#1}
\csname url@samestyle\endcsname
\providecommand{\newblock}{\relax}
\providecommand{\bibinfo}[2]{#2}
\providecommand{\BIBentrySTDinterwordspacing}{\spaceskip=0pt\relax}
\providecommand{\BIBentryALTinterwordstretchfactor}{4}
\providecommand{\BIBentryALTinterwordspacing}{\spaceskip=\fontdimen2\font plus
\BIBentryALTinterwordstretchfactor\fontdimen3\font minus
  \fontdimen4\font\relax}
\providecommand{\BIBforeignlanguage}[2]{{%
\expandafter\ifx\csname l@#1\endcsname\relax
\typeout{** WARNING: IEEEtran.bst: No hyphenation pattern has been}%
\typeout{** loaded for the language `#1'. Using the pattern for}%
\typeout{** the default language instead.}%
\else
\language=\csname l@#1\endcsname
\fi
#2}}
\providecommand{\BIBdecl}{\relax}
\BIBdecl

\bibitem{icassp}
W.~Cui and W.~Yu, ``Scalable reinforcement learning for routing in ad-hoc
  networks based on physical-layer attributes,'' in \emph{IEEE Int. Conf.
  Acoust., Speech, Signal Process. (ICASSP)}, Jun. 2021.

\bibitem{DSDV}
C.~E. Perkins and P.~Bhagwat, ``Highly dynamic destination-sequenced
  distance-vector routing {(DSDV)} for mobile computers,'' \emph{Proc. SIGCOMM
  '94}, pp. 234--244, Oct. 1994.

\bibitem{aodv}
C.~E. Perkins and E.~M. Royer, ``Ad-hoc on-demand distance vector routing,'' in
  \emph{IEEE Workshop Mobile Comput. Syst. Appl. (WMCSA)}, Feb. 1999.

\bibitem{tora}
V.~D. Park and M.~S. Corson, ``A highly adaptive distributed routing algorithm
  for mobile wireless networks,'' in \emph{IEEE Int. Conf. Comput. Commun.
  (INFOCOM)}, Apr. 1997.

\bibitem{ssa}
R.~Dube, C.~D. Rais, K.~Wang, and S.~K. Tripathi, ``Signal stability-based
  adaptive routing ({SSA}) for ad hoc mobile networks,'' \emph{IEEE Pers.
  Commun.}, vol.~4, no.~1, pp. 36--45, Feb. 1997.

\bibitem{aqor}
Q.~Xue and A.~Ganz, ``Ad hoc {Q}o{S} on-demand routing ({AQOR}) in mobile ad
  hoc networks,'' \emph{J. Parallel Distrib. Comput.}, vol.~63, no.~2, pp.
  154--165, Feb. 2003.

\bibitem{ssbr}
Y.~Taj and K.~Faez, ``Signal strength based reliability: A novel routing metric
  in {MANET}s,'' in \emph{Int. Conf. Netw. Secur. Wireless Commun. Trusted
  Comput.}, Apr. 2010.

\bibitem{gafni}
E.~Gafni and D.~Bertsekas, ``Distributed algorithms for generating loop-free
  routes in networks with frequently changing topology,'' \emph{IEEE Trans.
  Commun.}, pp. 11--18, Jan. 1981.

\bibitem{murthy}
S.~Murthy and J.~J. Garcia-Luna-Aceves, ``An efficient routing protocol for
  wireless networks,'' \emph{Mobile Netw. Appl.}, vol.~1, pp. 183--197, Jun.
  1996.

\bibitem{spine}
R.~Sivakumar, B.~S. Das, and V.~Bharghavan, ``Spine routing in ad hoc
  networks,'' \emph{Cluster Comput.}, vol.~1, pp. 237--248, Jun. 1998.

\bibitem{shiguang}
S.~Chen and K.~Nahrstedt, ``Distributed quality-of-service routing in ad hoc
  networks,'' \emph{IEEE J. Sel. Area Comm.}, vol.~17, no.~8, 1999.

\bibitem{markov}
R.~Bellman, ``A markovian decision process,'' \emph{J. Math. Mech.}, vol.~6,
  no.~5, pp. 679--684, 1957.

\bibitem{qrouting}
J.~A. Boyan and M.~L. Littman, ``Packet routing in dynamically changing
  networks: A reinforcement learning approach,'' \emph{Adv. Neural Inf.
  Process. Syst.}, vol.~6, Oct. 1999.

\bibitem{choi}
S.~Choi and D.~Yeung, ``Predictive q-routing: A memory-based reinforcement
  learning approach to adaptive traffic control,'' \emph{Adv. Neural Inf.
  Process. Syst.}, vol.~8, Dec. 1999.

\bibitem{chang}
Y.~Chang, T.~Ho, and L.~Kaelbling, ``Mobilized ad-hoc networks: a reinforcement
  learning approach,'' in \emph{Int. Conf. Auton. Comput. (ICAC)}, May 2004.

\bibitem{forster}
A.~Forster and A.~L. Murphy, ``A feedback-enhanced learning approach for
  routing in {WSN},'' in \emph{Commun. Distrib. Syst.}, Apr. 2007.

\bibitem{santhi}
G.~Santhi, A.~Nachiappan, M.~Z. Ibrahime, R.~Raghunadhane, and M.~K. Favas,
  ``Q-learning based adaptive {Q}o{S} routing protocol for {MANETs},'' in
  \emph{Int. Conf. Recent Trends Inf. Technol. (ICRTIT)}, Jun. 2011.

\bibitem{alharbi}
A.~Alharbi, A.~Al-Dhalaan, and M.~Al-Rodhaan, ``Q-routing in cognitive packet
  network routing protocol for {MANETs},'' \emph{NCTA 2014 - Proc. Int. Conf.
  Neural Comput. Theory Appl.}, pp. 234--243, Jan. 2014.

\bibitem{varun}
V.~K. Sharma, S.~S.~P. Shukla, and V.~Singh, ``A tailored {Q}-learning for
  routing in wireless sensor networks,'' in \emph{Int. Conf. Parallel, Distrib.
  Grid Comput.}, Dec. 2012.

\bibitem{leonardo}
L.~R.~S. Campos, R.~D. Oliveira, J.~D. Melo, and A.~D.~D. Neto,
  ``Overhead-controlled routing in {WSN}s with reinforcement learning,'' in
  \emph{Int. Conf. Intell. Data Eng. Autom. Learn. (IDEAL)}, 2012.

\bibitem{rahul}
R.~Desai and B.~P. Patil, ``Cooperative reinforcement learning approach for
  routing in ad hoc networks,'' in \emph{Int. Conf. Pervasive Comput. (ICPC)},
  Jan. 2015.

\bibitem{nurmi}
P.~Nurmi, ``Reinforcement learning for routing in ad hoc networks,'' in
  \emph{Int. Symp. Modeling Optim. Mobile, Ad Hoc Wireless Netw. Workshops
  (WiOpt)}, Apr. 2007.

\bibitem{qlearning}
C.~Watkins and P.~Dayan, ``Q-learning,'' \emph{Mach. Learn.}, vol.~8, pp.
  279--292, 1992.

\bibitem{DQN}
V.~Mnih, K.~Kavukcuoglu, D.~Silver, A.~Rusu, J.~Veness, M.~Bellemare,
  A.~Graves, M.~Riedmiller, A.~Fidjeland, G.~Ostrovski, S.~Petersen,
  C.~Beattie, A.~Sadik, I.~Antonoglou, H.~King, D.~Kumaran, D.~Wierstra,
  S.~Legg, and D.~Hassabis, ``Human-level control through deep reinforcement
  learning,'' \emph{Nature}, vol. 518, pp. 529--533, Feb. 2015.

\bibitem{DRL_survey}
N.~C. {Luong}, D.~T. {Hoang}, S.~{Gong}, D.~{Niyato}, P.~{Wang}, Y.~{Liang},
  and D.~I. {Kim}, ``Applications of deep reinforcement learning in
  communications and networking: A survey,'' \emph{IEEE Commun. Surveys Tut.},
  vol.~21, no.~4, pp. 3133--3174, 2019.

\bibitem{lspirouting}
P.~Wang and T.~Wang, ``Adaptive routing for sensor networks using reinforcement
  learning,'' in \emph{Int. Conf. Comput. Inf. Technol. (CIT)}, Sep. 2006.

\bibitem{lspi}
M.~G. Lagoudakis and R.~Parr, ``Model-free least squares policy iteration,'' in
  \emph{Neural Inf. Process. Syst. (NIPS)}, 2002.

\bibitem{dual}
W.~Yu and R.~Lui, ``Dual methods for nonconvex spectrum optimization of
  multicarrier systems,'' \emph{IEEE Trans. Commun.}, vol.~54, no.~7, pp. 1310
  -- 1322, Jul. 2006.

\bibitem{yiping}
Y.~Xing, R.~Chandramouli, S.~Mangold, and S.~{Shankar N}, ``Dynamic spectrum
  access in open spectrum wireless networks,'' \emph{IEEE J. Sel. Areas
  Commun.}, vol.~24, no.~3, pp. 626 -- 637, Mar. 2006.

\bibitem{songgao}
S.~Gao, L.~Qian, and D.~R. Vaman, ``Distributed energy efficient spectrum
  access in cognitive radio wireless ad hoc networks,'' \emph{IEEE Trans.
  Wireless Commun.}, vol.~8, no.~10, pp. 5202 -- 5213, Oct. 2009.

\bibitem{peiliang}
P.~Zuo, X.~Wang, W.~Linghu, R.~Sun, T.~Peng, and W.~Wang, ``Prediction-based
  spectrum access optimization in cognitive radio networks,'' in \emph{IEEE
  Int. Symp. Pers., Indoor Mobile Radio Commun. (PIMRC)}, Sep. 2018.

\bibitem{qingzhao}
Q.~Zhao, L.~Tong, A.~Swami, and Y.~Chen, ``Decentralized cognitive {MAC} for
  opportunistic spectrum access in ad hoc networks: A {POMDP} framework,''
  \emph{IEEE J. Sel. Areas Commun.}, vol.~25, no.~3, pp. 589--600, Apr. 2007.

\bibitem{mosleh}
S.~Mosleh, Y.~Ma, J.~D. Rezac, and J.~B. Coder, ``Dynamic spectrum access with
  reinforcement learning for unlicensed access in {5G} and beyond,'' in
  \emph{IEEE Veh. Technol. Conf. (VTC)}, May 2020.

\bibitem{yiding}
Y.~Yu, T.~Wang, and S.~C. Liew, ``Deep-reinforcement learning multiple access
  for heterogeneous wireless networks,'' \emph{IEEE J. Sel. Areas Commun.},
  vol.~37, no.~6, pp. 1277 -- 1290, Jun. 2019.

\bibitem{yingchang1}
L.~Zhang and Y.-C. Liang, ``Deep reinforcement learning for multi-agent power
  control in heterogeneous networks,'' \emph{IEEE Trans. Wireless Commun.},
  vol.~20, no.~4, pp. 2551 -- 2564, Apr. 2021.

\bibitem{yingchang2}
L.~Zhang, J.~Tan, Y.-C. Liang, G.~Feng, and D.~Niyato, ``Deep reinforcement
  learning based modulation and coding scheme selection in cognitive
  heterogeneous networks,'' \emph{IEEE Trans. Wireless Commun.}, vol.~18,
  no.~6, pp. 3281 -- 3294, 2019.

\bibitem{chunsheng}
C.~Xin, B.~Xie, and C.~Shen, ``A novel layered graph model for topology
  formation and routing in dynamic spectrum access networks,'' in \emph{IEEE
  Int. Symp. Dyn. Spectr. Access Netw. (DySPAN)}, Dec. 2005.

\bibitem{yongk}
Y.~Liu, L.~X. Cai, and X.~S. Shen, ``Spectrum-aware opportunistic routing in
  multi-hop cognitive radio networks,'' \emph{IEEE J. Sel. Area Comm.},
  vol.~30, no.~10, pp. 1958 -- 1968, Nov. 2012.

\bibitem{talay}
A.~C. Talay and D.~T. Altilar, ``Self adaptive routing for dynamic spectrum
  access in cognitive radio networks,'' \emph{J. Netw. Comput. Appl.}, vol.~36,
  no.~4, pp. 1140--1151, Jul. 2013.

\bibitem{sudeep}
S.~Tanwar, S.~Tyagi, N.~Kumar, and M.~S. Obaidat, ``{LA-MHR}: Learning automata
  based multilevel heterogeneous routing for opportunistic shared spectrum
  access to enhance lifetime of {WSN},'' \emph{IEEE Syst. J.}, vol.~13, no.~1,
  pp. 313--323, Mar. 2019.

\bibitem{bellman}
R.~Bellman, ``On the theory of dynamic programming,'' \emph{Proc. Natl. Acad.
  Sci. U.S.A.}, vol.~38, no.~8, pp. 716--719, 1952.

\bibitem{doubledqn}
H.~van Hasselt, A.~Guez, and D.~Silver, ``Deep reinforcement learning with
  double q-learning,'' in \emph{Proc. Assoc. Adv. Artif. Intell.}, Feb. 2016.

\bibitem{dueling}
Z.~Wang, T.~Schaul, M.~Hessel, H.~van Hasselt, M.~Lanctot, and N.~de~Freitas,
  ``Dueling network architectures for deep reinforcement learning,''
  \emph{Proc. Mach. Learn. Res.}, vol.~48, pp. 1995--2013, 2016.

\bibitem{epsilongreedy}
C.~Watkins, ``Learning from delayed rewards,'' Ph.D. dissertation, University
  of Cambridge, Jan. 1989.

\bibitem{optimalpayoff}
D.~H. Wolpert and K.~Tumer, ``Optimal payoff functions for members of
  collectives,'' \emph{Adv. Complex Syst.}, vol.~4, no.~2, pp. 265--279, 2001.

\bibitem{localreward}
Y.~Chang, T.~Ho, and L.~Kaelbling, ``All learning is local: Multi-agent
  learning in global reward games,'' in \emph{Adv. Neural Inf. Process. Syst.
  (NeurIPS)}, 2003.

\bibitem{idagent}
L.~Matignon, G.~J. Laurent, and N.~L. Fort-Piat, ``Independent reinforcement
  learners in cooperative {M}arkov games: a survey regarding coordination
  problems,'' \emph{Knowl. Eng. Review}, vol.~27, no.~1, pp. 1--31, 2012.

\bibitem{cui_spatial}
W.~Cui and W.~Yu, ``Spatial deep learning for wireless scheduling,'' \emph{IEEE
  J. Sel. Area Comm.}, vol.~37, no.~6, pp. 1248--1261, Jun. 2019.

\end{thebibliography}

\end{document}